\documentclass[twocolumn,usenatbib,useAMS]{mnras}

\PassOptionsToPackage{dvipsnames,svgnames,x11names}{xcolor}
\usepackage{xparse,soul}

\makeatletter
\NewDocumentCommand{\sotwo}{O{red}O{black}+m}
    {%
        \begingroup
        \color{#1}%
        \setul{-.5ex}{.4pt}%
        \def\SOUL@uleverysyllable{%
            \rlap{%
                \color{#2}\the\SOUL@syllable
                \SOUL@setkern\SOUL@charkern}%
            \SOUL@ulunderline{%
                \phantom{\the\SOUL@syllable}}%
        }%
        \ul{#3}%
        \endgroup
    }
\makeatother

\usepackage{graphicx}	
\usepackage{amsmath}	
\usepackage{amssymb}	
\usepackage{multicol} 
\usepackage{multirow}
\usepackage{bm}	
\usepackage{pdflscape}
\usepackage{caption}
\usepackage{subfigure}
\usepackage{siunitx}
\usepackage{color}

\newcommand{\kms}{\,km\,s$^{-1}$} 

\newcommand{\sbr}[1]{_{\textrm{#1}}}
\usepackage[T1]{fontenc}
\usepackage{ae,aecompl}

\usepackage{newtxtext,newtxmath}

\title[Universal profile of stacked filaments]{A Universal Profile for Stacked Filaments from Cold Dark Matter Simulations}

\author[T. Yang et. al.]{
Tianyi Yang$^{1}$\thanks{E-mail: T.Yang-24@sms.ed.ac.uk},
Michael J. Hudson$^{2,3,4}$\thanks{E-mail: mike.hudson@uwaterloo.ca},
Niayesh Afshordi$^{2,3,4}$\thanks{E-mail: nafshordi@pitp.ca}
\\
$^1$ Institute for Astronomy, University of Edinburgh, Royal Observatory, Blackford Hill, Edinburgh EH9 3HJ, UK\\
$^{2}$Department of Physics and Astronomy, University of Waterloo, Waterloo, ON, N2L 3G1, Canada\\
$^{3}$Waterloo Centre for Astrophysics, University of Waterloo, Waterloo, ON, N2L 3G1, Canada\\
$^{4}$Perimeter Institute of Theoretical Physics, 31 Caroline St. N., Waterloo, ON, N2L 2Y5, Canada
}

\pubyear{2022}

\begin{document}
\label{firstpage}
\pagerange{\pageref{firstpage}--\pageref{lastpage}}
\maketitle

\begin{abstract}
We study the stacked filaments connecting group-mass halo pairs, using dark-matter-only $N$-body simulations. We calculate the dark matter over-density profile of these stacked filaments at different redshifts as a function of the distance perpendicular to the filament axis. A four-parameter universal functional form, including three comoving scale radii and one amplitude parameter (core density), provides a good fit out to a radius of $20 h^{-1}\textrm{cMpc}$ for stacked filaments over a range of redshifts, lengths and masses. The scale radii are approximately independent of redshift but increase as power-laws with the comoving filament length. Lastly, we compare the scaling of the filament mass measured directly from the simulations to the predicted scaling from the halo-halo-matter three-point correlation function as a function of redshift and of the mass of the halo pairs. We find that both measured scalings are similar to, but somewhat shallower than the predictions, by 10\% and 30\%, respectively. These results provide a template to interpret present and upcoming observational results based on stacking, for example, weak lensing, thermal and kinetic Sunyaev-Zel'dovich, or X-ray observations.
\end{abstract}


\section{Introduction}\label{sec::intro}

On large scales, structure in the Universe is described by the so-called ``cosmic web'' \citep{cosmic_web_bond}, in which high density peaks (nodes) are connected by filaments and walls. The structure of this network is due to peaks and tidal fields present in the primordial matter density fluctuation field at the early times, sharpened in the non-linear regime as it collapses via gravitational instability \citep{zel'dovich_cosmic_web_formation} and forms the web-like pattern observed today. In the cosmic web, matter flows from underdense regions to higher density region, and ultimately, accumulates at highest density peaks. Filaments act as bridges that transport matter into these nodes.

Observationally, the presence of supercluster-scale filamentary structures, with sizes of order 100 comoving Mpc/$h$ (hereafter cMpc), has been established since the time of large galaxy redshift surveys \citep{perseus_filament,oort_fil_paper,deLapparentGellerHuchra1986, SDSS_great_wall} and detailed study of these and later redshifts surveys showed that filaments as traced by galaxies exist on smaller scales as well. Other observational evidence of filaments have been reported using X-ray emission \citep[e.g.][]{Fujita_x_ray_fil, Werner_X-ray_fil,Eckert_x_ray_fil, Chandra_fil, x_ray_stacked_paper,Vernstrom_X_ray_radio_stacking}, where most of these studies focus on relatively short and dense filaments between massive cluster pairs. 

The distribution of dark matter in filaments is a challenging problem. From a theoretical perspective, cosmological simulations allow us to identify filaments and study the matter distribution and its temporal evolution in more detail. \citet{Intercluster_filaments_in_Universe} identified individual filaments by extracting the dark matter particles between clusters. They found straight short filaments between haloes with $M_{\textrm{halo}}>10^{14}M_{\odot}$ and then averaged these to study their dark matter distribution, finding that the dark matter density profile should fall as $r^{-2}$ at large radii.

One observational probe of dark matter in filaments is weak gravitational lensing (WL). WL has been used to study the total matter distribution of individual filaments connecting rich clusters \citep{Dietrich_filament_detection, Jausac_MACS_filament}. However, the WL signal coming from individual filaments is weak, particularly if one wants to probe lower-mass filaments. 

The solution to the problem of low signal-to-noise is to identify and stack many filaments and measure the average WL signal from the stacked filament. To do so, one needs to first identify filaments.
Whereas dark matter haloes are well-defined, the morphology of filaments, as well as their multi-scale and diffuse nature makes the identification of these structures complicated and difficult. Due to this spatial complexity, a wide variety of sophisticated filament-finding algorithms have been proposed to find these filamentary structures in both simulations and observations \citep[see][for a detailed discussion]{Libeskind_2018,Rost_2019}. However, these filament identification algorithms depend on the assumption of several parameters, such as node properties, filament morphology, density threshold, etc., which can complicate the interpretation of filament properties. 

From an observational perspective, if one has a dense sample of galaxies then the above filament-finding methods can be applied, but, in practice, these methods are difficult to apply to realistic observational samples at intermediate ($z \sim 0.5$) redshifts, where spectroscopic data are presently sparse and exist only for bright galaxies such as luminous red galaxies (LRGs). Therefore, WL studies have focused on identifying galaxy pairs and stacking the WL signal from the filaments that are expected to be between them. Such a signal between LRGs separated by 5--20 cMpc/$h$ has now been detected by a number of authors \citep{clampitt_fil_detection, Seth_filament_paper, He_fil_paper,  new_filament_paper, Kondo_fil_paper, Tianyi_filament_paper}. 

It is important to note that the WL measurements described above are not simply of the total mass between a pair of haloes, but rather the excess mass over and above that expected from summing the correlated mass that would be associated with two isolated haloes of the same mass. In this sense, this measurement is known as the {\it connected part} of the three-point halo-halo-mass correlation function.

The goal of this paper is to study the mass and density structure of such stacked N-body filaments as a function of the length of the filament, its redshift and the mass of the haloes that define its ends. The paper is organized as follows. In Section \ref{sec::simulation_usage}, we describe the simulation and halo catalogue used for filament identification. In Section \ref{sec::fil_identify}, we introduce the universal over-density profile (Section \ref{ssec:Density_profile_and_scaling_relationship_of_filaments}) and fit this form to the stacked profile of filaments defined only in 3D distance space in the simulation box (Section \ref{ssec::data_selection_galaxy_pairs}). Section \ref{sec::Density_profile_and scaling relationship_of_filaments_with_selection_biases} focuses on the effect of various selection criteria, as chosen by different observational studies, on the resulting filament properties, and we summarise our results and conclude in Section \ref{sec::discussion} and Section \ref{sec::conclusion}. As mentioned above, unless specified, the mass of filament in this paper is defined as the excess mass over the matter distribution from the superposition of two isolated haloes.  

\section{The Simulation}\label{sec::simulation_usage}

We use the MultiDark Planck 2 simulation \citep{MDPL2_ref, MDPL2_database_ref}, a dark-matter-only simulation with mass resolution of $m_{\textrm{DM}} = 1.5\times10^{9} M_{\odot}/h$ in a periodic cube of size $1,000\,\textrm{cMpc}/h$ evolved with a $\textit{Planck}$-2013 \citep{planck2013} cosmology with parameters  $\Omega_{m} = 0.307, \Omega_{\Lambda} = 0.693, \Omega_{b} = 0.048, H_{0} = 67.8 \rm ~km~s^{-1}~Mpc^{-1}, \sigma_{8} = 0.829$ and $n_{s} = 0.96$. Full particle catalogues\footnote{MDPL2 particle catalogues are available at \url{https://www.cosmosim.org/cms/files/simulation-data/}} are available for 4 different snapshots at at $z = 0.0$, 0.49, 1.03 and 2.53 respectively, and we use first three of these in this paper. 

Halo catalogues\footnote{\texttt{RockStar} halo catalogues are available at \url{https://www.cosmosim.org/cms/documentation/database-structure/tables/rockstar/}} are identified by the \texttt{RockStar} halo-finding algorithm \citep{rockstar_ref}. Halo positions and velocities are directly read from the database, but there are multiple definitions regarding the halo mass values. Halo masses are calculated within various thresholds, such as within a specific density threshold $\delta$ relative to the mean matter density or the critical density. In this study, we choose the ``virial'' mass defined within a spherical volume of radius $R\sbr{vir}$ with mean density $\rho_{\textrm{vir}}$, where $\rho_{\textrm{vir}}$ is $\delta_{\textrm{vir}}\rho_{\textrm{m}}$ at that redshift. The ``virial'' over-density threshold, denoted $\delta_{\textrm{vir}}(z)$, is given by the approximation of \citet{virial_overdensity_ref}, which yields 337 at $z = 0.0$. 

In this study, only haloes with no more-massive haloes within the halo radii are selected (so-called distinct haloes). These haloes are flagged with \texttt{pID = -1} in the \texttt{RockStar} database. Due to the completeness and resolution of simulation data, we only select distinct haloes with virial mass $M_{\textrm{vir}}\geqslant 10^{11} ~h^{-1}M_{\odot}$. This gives the fraction of particles bound to haloes of this mass or greater equal to 0.40 at $z = 0.0$, 0.32 at $z = 0.49$ and 0.24 at $z = 1.03$ respectively. The mean virial masses of the haloes are: $3.59\times10^{13}h^{-1} M_{\odot}$ at $z = 0.0$, $2.85\times10^{13}h^{-1} M_{\odot}$ at $z = 0.49$, and $2.28\times10^{13}h^{-1} M_{\odot}$ at $z = 1.03$. 

\section{Filament Identification}\label{sec::fil_identify}

Using the simulation data, filaments are defined as the particles between halo-halo pairs that are close in 3D coordinates. Section \ref{ssec::data_selection_galaxy_pairs} discusses the halo-halo pair construction as well as the procedure for isolating filament over-density signal. Then we fit the stacked over-density profile using the universal functional form proposed in Section \ref{ssec:Density_profile_and_scaling_relationship_of_filaments} and discuss how the best fitting parameters vary with redshift and filament size in Section \ref{ssec:Scaling_of_the_filament_density_parameters_no_selection_cut}.

\subsection{Physical and non-physical pairs}
\label{ssec::data_selection_galaxy_pairs}

\begin{figure}
\includegraphics[width=\columnwidth]{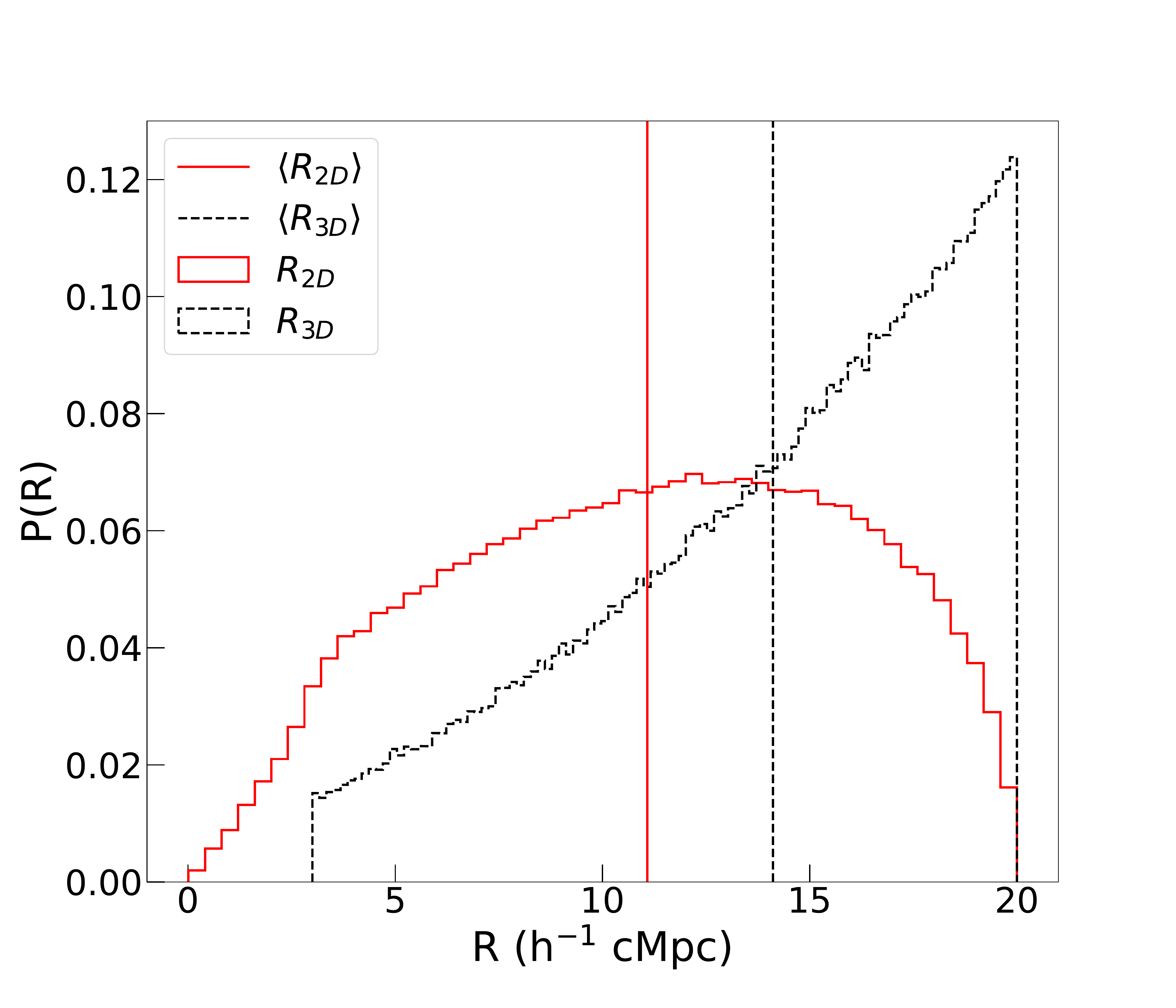}
   \caption{Probability distribution of pair separation in 2D (red solid line) and in 3D (black dashed line). Pairs are constructed by selecting haloes with 3D $R$ separation within $3h^{-1} \textrm{cMpc} \leqslant R_{\textrm{3D}} \leqslant 20 h^{-1} \textrm{cMpc}$. Average values of $R_{\textrm{sep}}$ in 2D and 3D are shown as vertical lines on the same plot. There are 335,560 pairs in total (using pairs selected at $z=0.0$ for illustration).}
   \label{R_sep_pairs_all}
\end{figure}

To find proxies for galaxy groups and clusters, we only identify filaments between massive haloes with virial mass greater than $10^{13}h^{-1} M_{\odot}$, which is similar to the halo mass range in which Luminous Red Galaxies (LRGs) reside \citep[justified by weak lensing and clustering analyses, e.g.][]{WL_LRG_halo_mass,clustering_anal_LRG_halo_mass}. Since group-sized haloes that are close in three-dimensional space are expected to have filaments connected, physical halo pairs are simply defined by the comoving separations of their haloes in 3D. We select all halo pairs with $R_{\textrm{3D}}$ between $3~h^{-1} \textrm{cMpc}$ and $20 ~h^{-1} \textrm{cMpc}$ at the three different simulation snapshots. To reduce computational time, at each snapshot, our parent catalogue is constructed by randomly selecting one-tenth of the total number of pairs. This yields 335,560 pairs at $z = 0.0$, 210,667 pairs at $z = 0.49$ and 84,006 pairs at  $z = 1.03$. The distributions of projected (2D) and 3D comoving separations between halo pairs are shown in the left panel of Figure \ref{R_sep_pairs_all}, where the mean values of two distances over all filaments in the sample are shown as vertical lines on the same plot. For illustration, only pairs selected at $z = 0.0$ are shown here.

To isolate over-density of the filament properly, we remove all particles within a spherical volume of $2 R_{\textrm{vir}}$ around each halo in order to exclude mass associated with the haloes at either end of the filament. To calculate the excess density, $\Delta \rho$,  associated with the filament over and above what would be expected from the sum of two isolated haloes of the same virial mass, we construct a mock sample of halo pairs that are not physically connected in 3D. To achieve this, for each paired halo, we select another unpaired halo in the field with matched virial mass (without replacement), and then translate all particles in the simulation (centered at the selected unpaired halo) to the position of the paired halo with a random orientation defining the filament axis. We then repeat this process for the other halo in the pair, and then repeat for all halo pairs. This way allows us to obtain a sample of \emph{non-physical} pairs whose virial masses are the same those of the true physical halo pairs but whose density distribution represents the sum of two two-point halo-matter cross-correlations. The filament is then defined as the excess density that remains after the subtracting the density profile of the non-physical pairs from that of the physical pairs. Since non-physical pairs are constructed by a superposition of two simulation boxes, the background needs to be subtracted twice for non-physical pairs, while one only needs to do so once for the physical pairs. 

\begin{figure*}
    \begin{minipage}[htb!]{\textwidth}
        \centering
        \includegraphics[width=\linewidth]{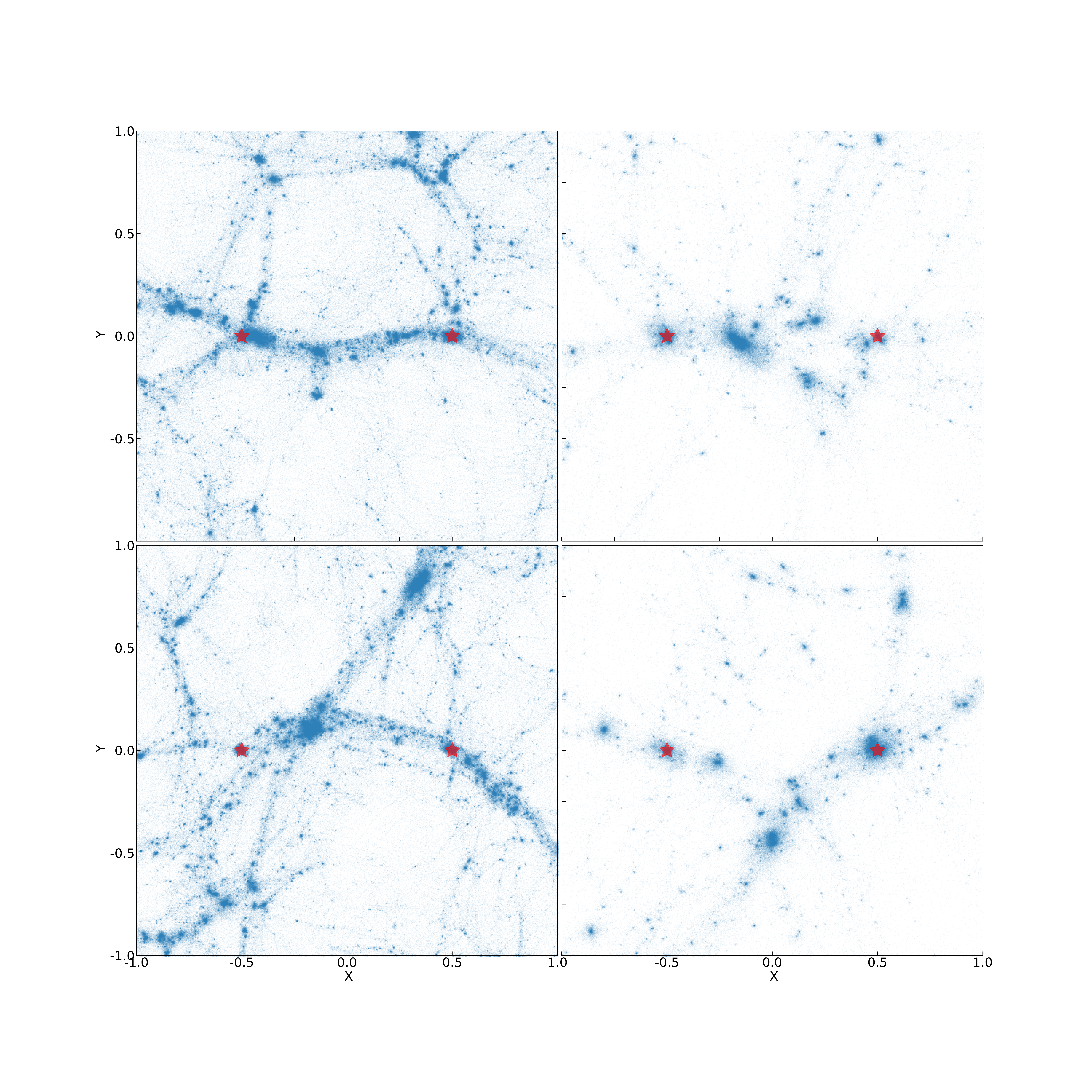}
    \end{minipage}%
    \vskip-0.6in
\caption{The dark matter particle distribution between four physical pairs. Pairs are selected with $3h^{-1} \textrm{cMpc} \leqslant R_{\textrm{3D}} \leqslant 20 h^{-1} \textrm{cMpc}$ at snapshot $z = 0.49$. For each pair, all particles within a slice of $\pm r_{\rm z}$ (defined in Section \ref{ssec:Density_profile_and_scaling_relationship_of_filaments}) are projected onto $x-y$ plane. Coordinates are normalised such that the haloes at either end of the filament, shown as red stars, are positioned at (+0.5,0) and (-0.5,0).}
\label{physical_pairs_scatter_fig}
\end{figure*}

Our approach of identifying filaments is largely free from arbitrary parameter choices used in other filament-finding algorithms. The filaments are identified between haloes above a certain mass threshold separated by a given length. One caveat of this method is that there is no guarantee that there will be a straight filament, or any filament at all, between any given halo pair. This can be seen from Figure \ref{physical_pairs_scatter_fig}, where we demonstrate the DM particle distribution between four individual physical pairs selected at $z = 0.49$. To generate this plot, we project all particles within a slice of $\pm r_{\rm z}$ (defined in Section \ref{ssec:Density_profile_and_scaling_relationship_of_filaments}) in front of/behind the filament $xy$ plane. The filament axis is rotated to align with the horizontal ($x$) axis of the plot, with the coordinates normalised by the halo pair separation, $R_{\textrm{3D}}$, such that the two halo ``ends'' are positioned at $(x,y) = (+0.5,0)$ and $(-0.5,0)$. The DM particle field is rotated in the same way and then projected into the $(x,y)$ plane. It can be seen from these examples that, indeed, filamentary structures with different morphologies  are included in our catalogue. However, by stacking a large population of filament members, we recover an average stacked filament, which has a different structure than any individual filament.

\begin{figure}
\includegraphics[width=\columnwidth]{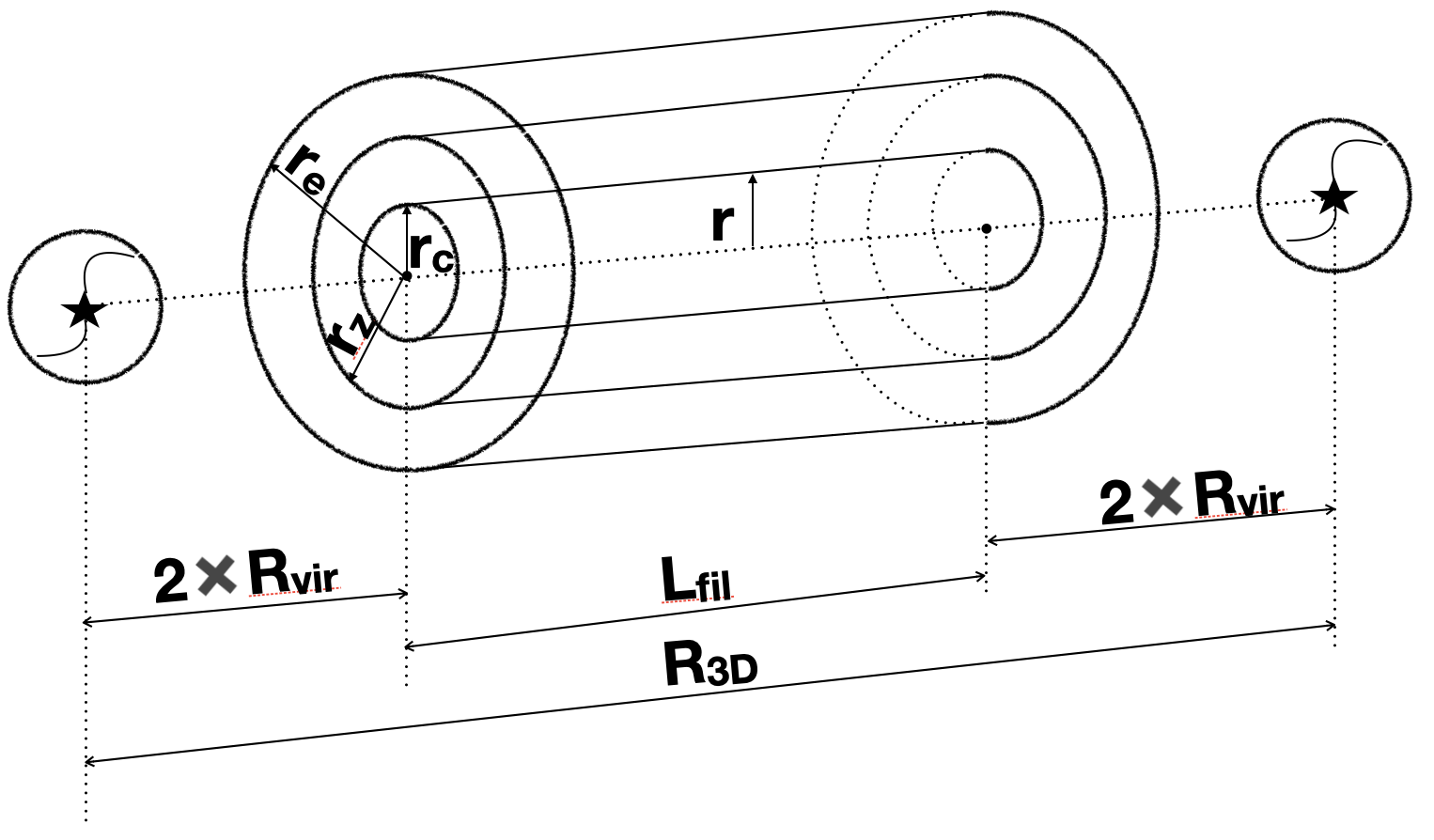}
   \caption{The geometry of the stacked filament model in our study, where only a cylindrical region with length of $L_{\rm fil} \equiv R_{\textrm{3D}} - 2R_{\textrm{vir, halo 1}} - 2R_{\textrm{vir, halo 2}}$ is considered.  The core radius, $r\sbr{c}$, the zero-crossing radius,  $r_{\textrm{z}}$, and the ``environmental'' radius, $r_{\textrm{e}}$, are three comoving scale-radii appearing in the excess density profile (Figure \ref{params_fit_excluding_ends_all_fit}) as discussed in Section \ref{ssec:Density_profile_and_scaling_relationship_of_filaments}.}
   \label{fil_model_sketch}
\end{figure}

We then compute the density profile of each filament. It is calculated within a cylinder centred on the line connecting the two halos with a cylindrical comoving radius $r < 20~h^{-1} \textrm{cMpc}$ and a filament length defined as $L_{\textrm{fil}} \equiv R_{\textrm{3D}} - 2R_{\textrm{vir, halo 1}} - 2R_{\textrm{vir, halo 2}}$. The geometry is shown in Figure \ref{fil_model_sketch}. The density profile is computed by summing up all DM particle mass in cylindrical shells and dividing by the shell volumes and subtracting the mean density. Then the density profile computed from individual filament are stacked together to obtain an average over-density profile. While measuring the over-density field within each filament cylinder, we consider two different possibilities: counting all particles or counting only particles that are within the virial radii of haloes with $M_{\textrm{vir}}\geqslant 10^{11} ~h^{-1}M_{\odot}$.

\begin{figure}
\includegraphics[width=\columnwidth]{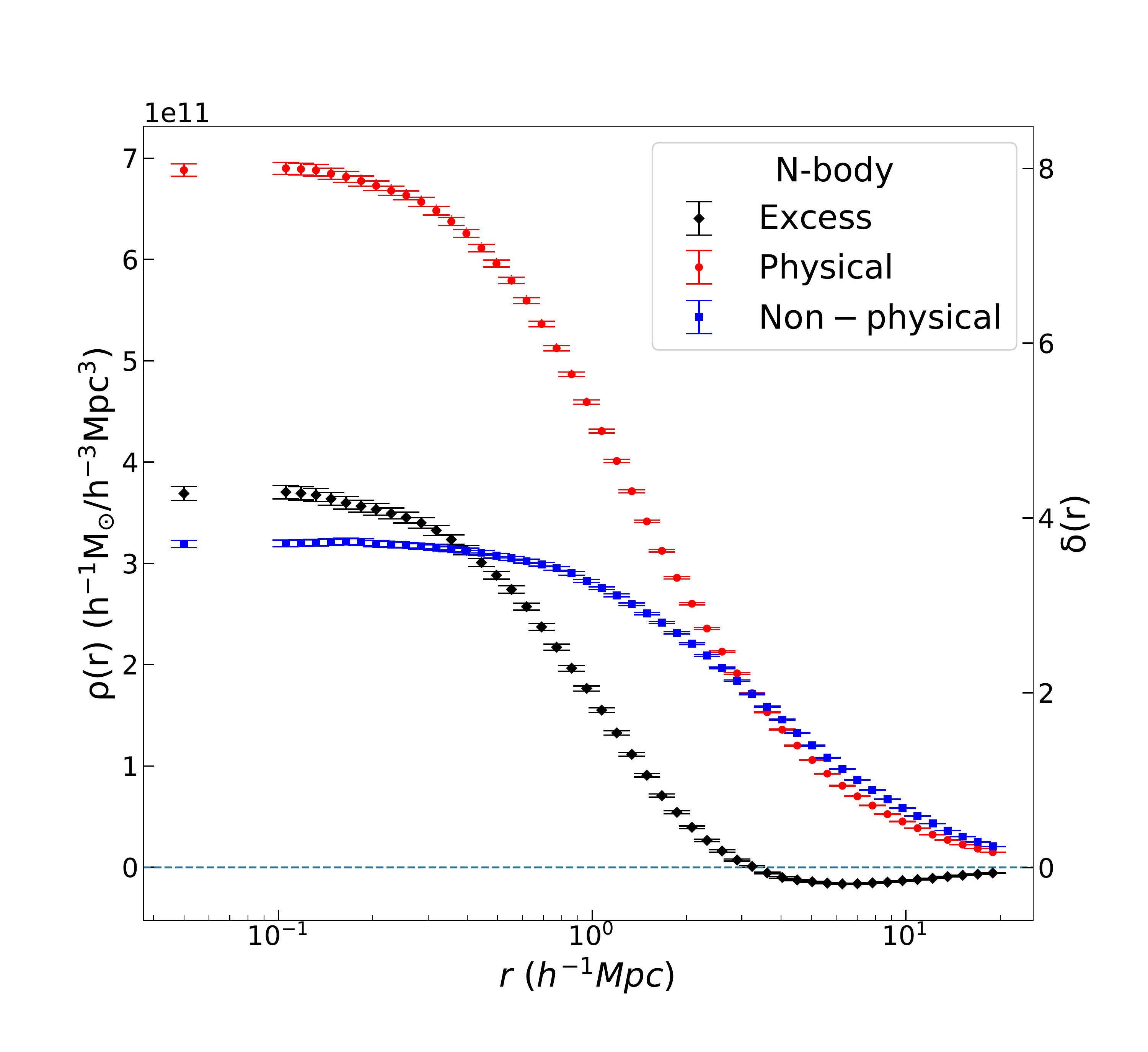}
   \caption{The profile of the mass density as a function of distance perpendicular to the cylinder axis for all particles (as defined in Section \ref{ssec::data_selection_galaxy_pairs}). The profile is obtained by averaging filament pairs with $3h^{-1} \textrm{cMpc} \leqslant R_{\textrm{3D}} \leqslant 20 h^{-1} \textrm{cMpc}$ at snapshot $z = 0.0$. There are 335,560 pairs in total. Standard deviations are obtained from the square root of diagonal elements in covariance matrix. The red circles show the density profile of physical pairs after subtracting the background density. The blue squares show the density profile of the non-physical pairs after subtracting the background twice. Finally, the black diamonds show the excess density profile after subtracting the non-physical profile from the physical one.}
   \label{particle_profile_total_nbody_z00}
\end{figure}

Figure \ref{particle_profile_total_nbody_z00} shows the stacked density profile at $z = 0.0$ for physical and non-physical filament after subtracting the background density (once for physical pairs, twice for non-physical pairs), counting all particles. The difference of these two curves, which is the excess signal, denoted as the \textit{connected part} in this study, is shown as black diamond points on the same figure for comparison. The excess density profile is computed as $\Delta \rho(r) \equiv \rho(r)\sbr{physical}-\rho(r)\sbr{non-physical}$. To better characterise the excess density profile after the subtraction, we propose a functional form to fit the measured data points from N-body simulation, which will be discussed in the next section.

\subsection{Density profile and scaling relationship of filaments}\label{ssec:Density_profile_and_scaling_relationship_of_filaments}

\begin{figure*}
    \begin{minipage}[htb!]{1.0\textwidth}
        \centering
        \includegraphics[width=1.0\linewidth]{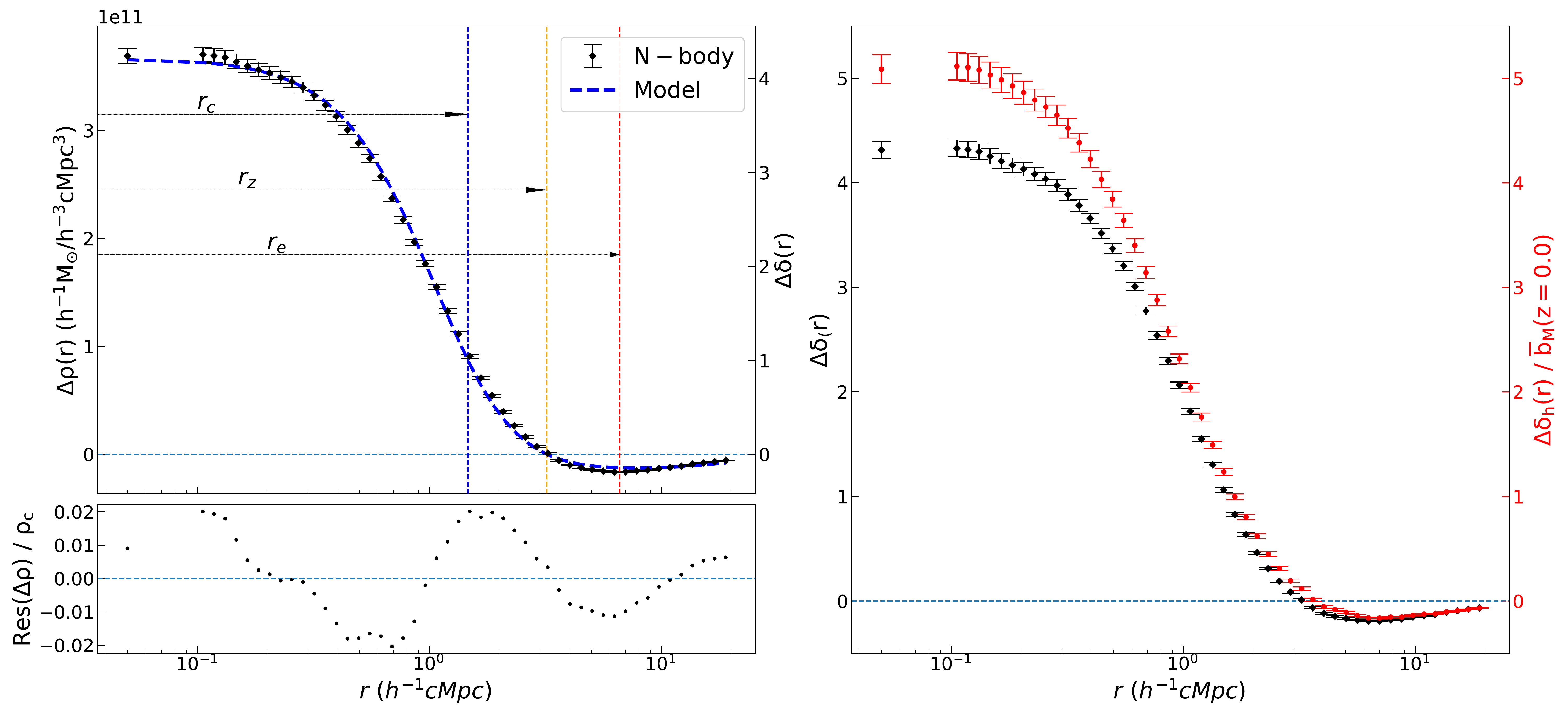}
    \end{minipage}%
\caption{Left panel: the profile of the excess mass density and over-density contrast as a function of distance perpendicular to the cylinder axis for all particles. The profile is obtained by averaging filament pairs with $3h^{-1} \textrm{cMpc} \leqslant R_{\textrm{3D}} \leqslant 20 h^{-1} \textrm{cMpc}$ at snapshot $z = 0.0$. There are 335,560 pairs in total. The best fit and standard deviation of the points are over-plotted on the same figure. The residual between the N-body measurements and the best fitting line is shown in the bottom panel, where values on the $y$ axis are normalised by the best fitting core density $\rho_{\textrm{c}}$. Right panel: a comparison of the computed over-density contrast profile for all particles ($\Delta \delta(r)$, black diamonds) and halo particles only ($\Delta \delta_{\textrm{h}}(r)$, red circles), where $\Delta \delta_{\textrm{h}}(r)$ is normalised by the mass-weighted halo bias measured for whole simulation at $z = 0.0$ (see Equation \ref{eqn::bias_cal}).}
\label{params_fit_excluding_ends_all_fit}
\end{figure*}

Figure \ref{params_fit_excluding_ends_all_fit} shows the resulting stacked excess filament density profile selected at $z = 0.0$, considering  all particles. On the right-hand axis, we show the over-density contrast profile, which is computed as $\Delta \delta(r) \equiv \Delta \rho(r)/\rho_{\textrm{m, all particles}}$. The profile shows a ``core'' in the innermost region, and it keeps decreasing as a power law till a point where the density equals to zero. At larger radii, there is a negative valley, which is expected because the entire enclosed excess mass profile must integrate to $0$ when integrated over all space. In other words, the excess density at small radii must be compensated by a negative region elsewhere.  The profile appears to asymptote to zero at large radii, as expected. To characterise this behaviour, we propose the following density profile:
\begin{equation}\label{eqn::fil_density_profile}
    \Delta \rho(r) = \dfrac{{\rho}_\text{c}\left(1-\frac{r^4}{r_{\textrm{z}}^4}\right)}{\left(\frac{r^4}{r_{\textrm{e}}^4}+\frac{r^2}{r_{\textrm{c}}^2}+1\right)^2},
\end{equation}
where $r$ is the distance perpendicular to the filament axis, and the core density at the $r=0$ axis is given by $\rho_{\textrm{c}}$. To characterise the density profile, we include three different comoving scale radii in the proposed functional form. In the numerator, $r_{\textrm{z}}$ is the zero-crossing radius, i.e. the point where the density profile first goes negative. In the denominator, $r_{\textrm{c}}$ is the core radius, while $r_{\textrm{e}}$ is the ``environmental'' radius that characterises the size of the emptied region, or the``valley'' in the negative tail. The excess mass of the filament is then computed by integrating the above profile from zero to the best fitting $r_{\textrm{z}}$.

Data points shown in Figure \ref{params_fit_excluding_ends_all_fit} are then fitted with Equation \ref{eqn::fil_density_profile} and the residual values between the N-body measurements and best-fit curves, normalised by the best-fit $\rho_{\textrm{c}}$, are shown on the bottom panel. The error on each data point is given by the diagonal element of the covariance matrix. For the excess mass density, the best-fit parameters are $\langle \rho_{\textrm{c}} \rangle =3.668\times10^{11} ~h^{-1}M_{\odot}/h^{-3}\textrm{cMpc}^3$ (corresponds to a value of $\langle \delta_{\textrm{c}} \rangle = \langle \rho_{\textrm{c}} \rangle/\rho_{\textrm{m, all particles}} = 4.290$), where $\langle ... \rangle$ denotes an average of over all filaments in the sample. And $\langle r_{\textrm{c}} \rangle =1.464~h^{-1}\textrm{cMpc}$, $\langle r_{\textrm{z}} \rangle = 3.208~h^{-1}\textrm{cMpc}$ and $\langle r_{\textrm{e}} \rangle =6.604~h^{-1}\textrm{cMpc}$. Considering only the mass density of particles in haloes, the best-fit parameters are $\langle \rho\sbr{c,h} \rangle =2.489\times10^{11} ~h^{-1}M_{\odot}/h^{-3}\textrm{cMpc}^3$ (corresponding to $\langle \delta\sbr{c,h} \rangle = \langle \rho\sbr{c,h} \rangle/\rho_{\textrm{m, halo particles only}} = 7.341$), $\langle r\sbr{c,h} \rangle =1.409~h^{-1}\textrm{cMpc}$, $\langle r\sbr{z,h} \rangle = 3.768~h^{-1}\textrm{cMpc}$ but $\langle r\sbr{e,h} \rangle$ is not well-constrained. With the best fitting density profile, we then compute the filament mass within the cylinder by integrating the density profile up to $\langle r_{\textrm{z}} \rangle$. This gives $\langle \Delta M_{\textrm{fil}} \rangle$ = $2.109\times10^{13} ~h^{-1}M_{\odot}$ (all particles) and $1.449\times10^{13} ~h^{-1}M_{\odot}$ considering  halo particles only.

\begin{figure*}
    \begin{minipage}[htb!]{\textwidth}
        \centering
        \includegraphics[width=\linewidth]{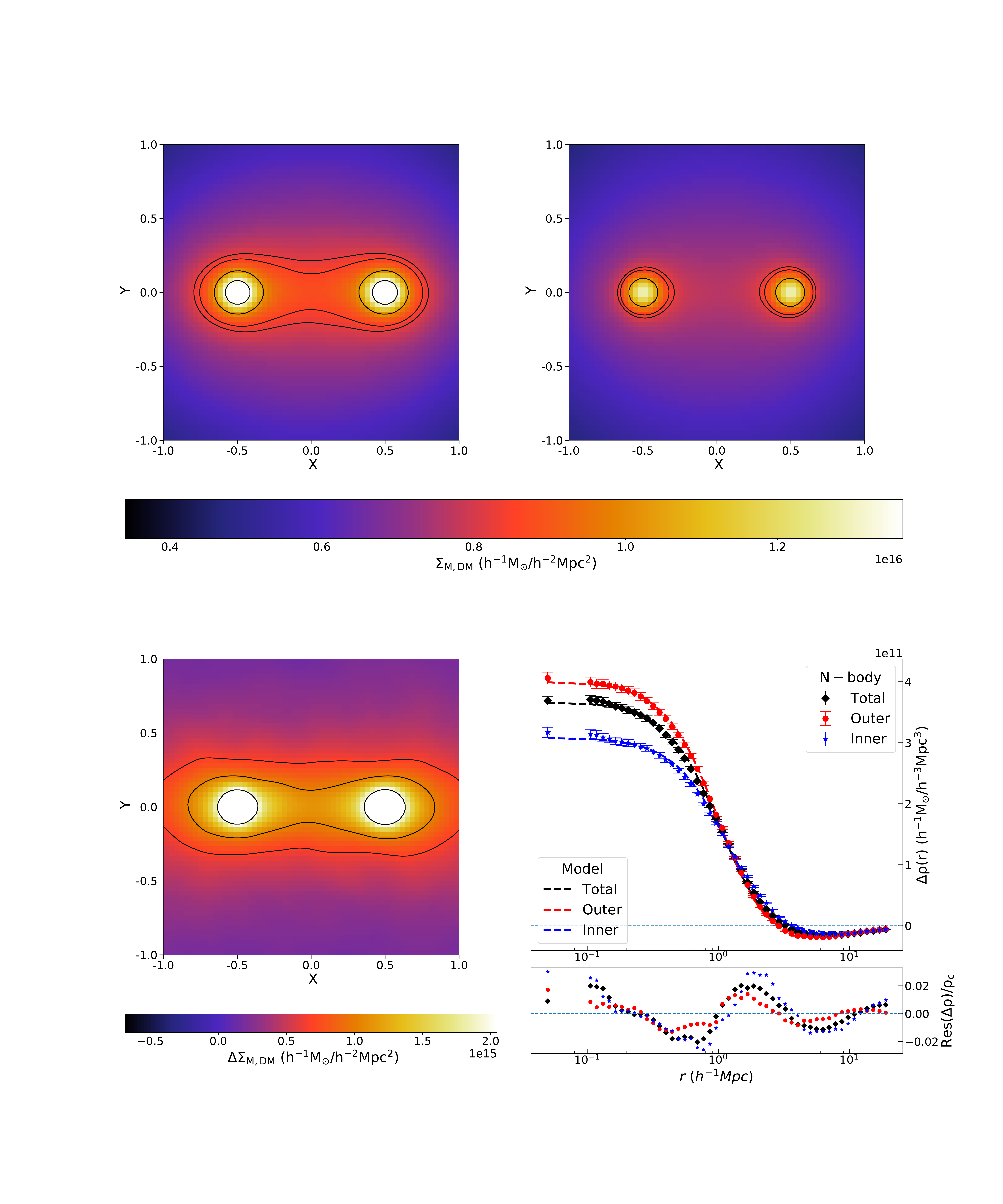}
    \end{minipage}%
    \vskip-0.8in
\caption{Dark matter surface mass density map for the stacked physical pairs (\textit{upper left}), the stacked non-physical pairs (\textit{upper right}) and the excess map obtained from the subtraction of two maps (\textit{lower left}). Note the different color range for the excess map compared with the upper panels. One unit in the rescaled coordinate system corresponds to approximately 14 $h^{-1}$ cMpc. A ``dumbbell'' structure is noticeable between physical pairs and in the excess. On the lower right panel, we demonstrate the profile of the excess mass density as a function of distance perpendicular to the cylinder axis at different filament segments. Stacked profiles obtained from different parts of filament are differentiated by colors (black diamonds: averaging over all filament regions; red circles: averaging over the outer regions only; blue stars: averaging over the inner regions only; see text for details). The residual between the N-body measurements and the best fitting line is shown in the bottom plot of the panel, where values on the $y$ axis are normalised by the best fitting core density $\rho_{\rm c}$.}
\label{projected_num_dens_map}
\end{figure*}

Our methodology might select pairs where there are no DM particles between them, and this is the reason why it is vital to stack a large population of pairs to obtain a significant detection of filaments between haloes. Figure \ref{projected_num_dens_map} shows the DM particle surface mass density map between pairs, which is obtained by stacking all individual halo pairs as demonstrated in Figure \ref{physical_pairs_scatter_fig}. The ``dumbell'' structure is present between stacked physical pairs as well as in the excess map. Also, filaments are denser close to the haloes at the two ends compared to the middle of the filament.  This is shown in the lower right panel of Figure \ref{projected_num_dens_map}: for each filament, we divide the filament cylinder into four segments with equal length along the axis, where the outer two segments characterises the profile close to the halo ends, while the inner two parts reflect the matter distribution in the middle of filaments. Then we take an average of two outer segments and two inner segments respectively\footnote{Stacked profiles of individual outer(inner) segments are close due to symmetry.}. The best fitting curves for different segments are overplotted on the same figure. This density dependence along the filament axis has been illustrated by many previous studies with different components in stacked filaments, such as \citet{Intercluster_filaments_in_Universe} for dark matter, \citet{Martinez_2016_galaxy} for galaxies and \citet{Rost_2021_the300} for both dark matter and gas. Our finding here is consistent with these works.

\subsection{Scaling of the filament density parameters}\label{ssec:Scaling_of_the_filament_density_parameters_no_selection_cut}

\begin{figure*}
    \begin{minipage}[htb!]{\textwidth}
        \centering
        \includegraphics[width=\linewidth]{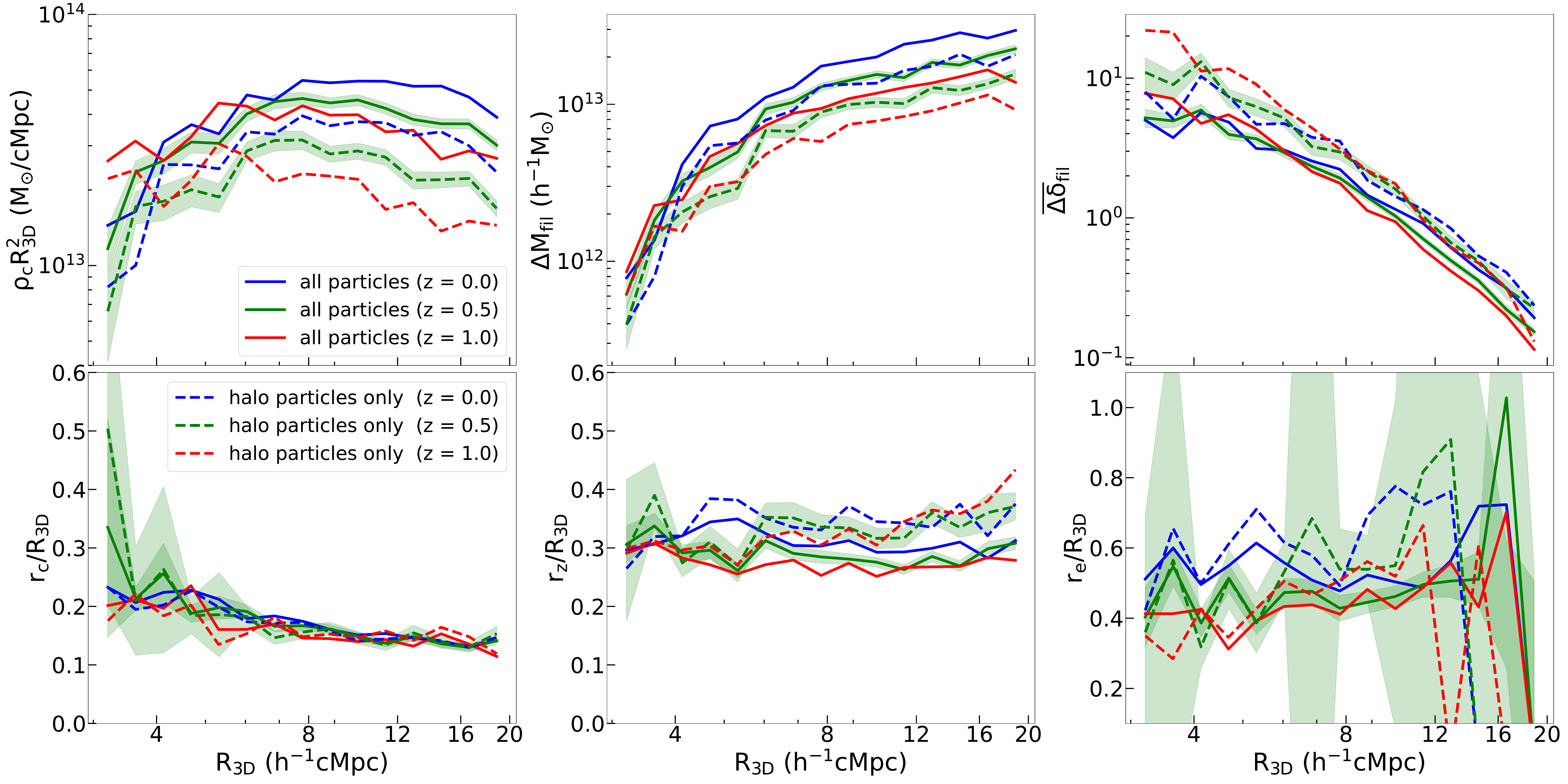}
    \end{minipage}%
\caption{The scaling of the best fitting density profile parameters, including core density $\rho_{\textrm{c}}$ (upper left), filament mass integrated from the excess density profile (Equation \ref{eqn::fil_density_profile}) within filament cylinder $\Delta M_{\textrm{fil}}$ (upper middle), mean over-density of the filament $\overline{\Delta \delta}_{\textrm{fil}}$ (upper right) and scaling radii $r_{\textrm{c}}, r_{\textrm{z}}, r_{\textrm{e}}$ (bottom row), as a function of the separation between the haloes, $R_{\textrm{3D}}$, for the selection-free case. Solid lines show the best fit for all particles and dashed lines shows the best fit considering halo particles only. On each panel, redshift evolution of different quantities are distinguished by different colors. Shaded regions show the error in the mean around the best fitting parameters. Only uncertainties for filaments at $z=0.49$ are displayed here for illustration. Best fitting values and uncertainties for different density profile parameters are given
in Table \ref{table_params_summary_selection_free} for reference.}
\label{rsep3d_binned_free_select}
\end{figure*}

It is interesting to study how the parameters of the density profile depend on the separation and redshift of the halo pair. Figure \ref{rsep3d_binned_free_select} shows the resulting best-fit parameters as a function of the separation of the halo pair, $R_{\textrm{3D}}$, and redshift, both considering all particles, and considering the particles in haloes only\footnote{Unless specified, all results presented in the following figures are for all particles.}. The shaded regions show the uncertainties in the best fitting parameters as determined by bootstrap resampling as follows. For all pairs within a specific $R_{\textrm{3D}}$ bin, we construct a bootstrap catalogue by resampling halo pairs with replacement but keeping the sample size identical as the original. We then repeat this process 1000 times and compute the average as well as the standard deviation across the bootstrap samples. Best fitting values for different density profile parameters are tabulated in Table \ref{table_params_summary_selection_free}. The top left panel shows the values of core density $\rho_{\textrm{c}}$ as a function of $R_{\textrm{3D}}$. For both cases, it can be seen that at small separation, the core density of filaments increases rapidly as $R_{\textrm{3D}}$, but for filaments with large enough 3D separations (when $R_{\textrm{3D}}\gtrsim6~h^{-1}\textrm{cMpc}$), $\rho_{\textrm{c}}$ roughly goes as $R_{\textrm{3D}}^{-2}$. Interestingly, as shown in the bottom row, the radii, especially $r_{\textrm{c}}$ and $r_{\textrm{z}}$, when normalised by $R_{\textrm{3D}}$, do not significantly depend on redshift, nor, in the case of $r_{\textrm{z}}$ on $R_{\textrm{3D}}$. A good approximation is $r_{\textrm{z}} = 0.3R\sbr{3D}$ and $r_{\textrm{c}}/R\sbr{3D} = -0.14\times\textrm{log}_{10}(R\sbr{3D})+0.30$ for curves considering all particles.

The rightmost panel of top row of Figure.\ \ref{rsep3d_binned_free_select} shows the mean excess density of the filament, $\overline{\Delta \delta}\sbr{fil}$, which, like the filament mass, is computed within a cylinder the radius of which is the best fitting $r_{\textrm{z}}$, and the length which is defined as the comoving distance between two haloes but excluding 2$R_{\textrm{vir}}$ around each halo. The background density at each redshift is determined either using all particles in the simulation box, or just the particles in haloes with $M_{\textrm{vir}}\geqslant 10^{11} ~h^{-1}M_{\odot}$. The mean over-density, $\overline{\Delta \delta}\sbr{fil}$, is a decreasing function of $R_{\textrm{3D}}$, with the shortest filaments having mean over-densities $\sim 8$, but filaments with $R_{\textrm{3D}} \gtrsim 9 ~h^{-1} \textrm{cMpc}$ are only mildly non-linear, with over-densities $\lesssim 1$.

It is interesting to compare the relationship between $\overline{\Delta \delta}_{\textrm{fil, halo particles only}}$ and $\overline{\Delta \delta}_{\textrm{fil, all particle}}$ on the one hand with the (mass-weighted) linear halo bias in the simulation as a whole on the other. The predicted mass-weighted mean halo bias can be computed at three different redshifts using
\begin{equation}\label{eqn::bias_cal}
    \overline{b}_{M}(z) = \frac{\int_{10^{11} ~h^{-1}M_{\odot}}^{+\infty}\textrm{d}M\frac{\textrm{d}n}{\textrm{d}\textrm{ln}M}(M,z)b(M,z)}{\int_{10^{11} ~h^{-1}M_{\odot}}^{+\infty}\textrm{d}M\frac{\textrm{d}n}{\textrm{d}\textrm{ln}M}(M,z)},
\end{equation}
where $M$ is the virial mass defined in the same way as in the simulation, the halo mass function is from \cite{halo_mass_function} and halo bias function is from \citep{Halo_bias_model}. Results are $\overline{b}_{M}(z = 0.0) = 1.45$, $\overline{b}_{M}(z = 0.49) =1.61$ and $\overline{b}_{M}(z = 1.03) =1.83$. This comparison is displayed in the right panel of Figure \ref{params_fit_excluding_ends_all_fit}. It can be seen that the amplitude of $\Delta \delta_{\textrm{h}}(r)$, which is the over-density contrast profile computed with halo particles only, overlaps with $\Delta \delta(r)$ at large scales after normalising the $\Delta \delta_{\textrm{h}}(r)$ profile by the linear bias $\overline{b}_{M}$. However, the rescaled $\Delta \delta_{\textrm{h}}(r)$ profile is slightly greater than $\Delta \delta(r)$ in the region with $r<r_{\textrm{z}}$. This suggests that haloes in the non-linear filaments are more biased than in the simulation as whole. 

\subsection{The mass function of haloes in filaments}\label{ssec:halo_mass_function}

\begin{figure*}
    \begin{minipage}[htb!]{\textwidth}
        \centering
        \includegraphics[width=\linewidth]{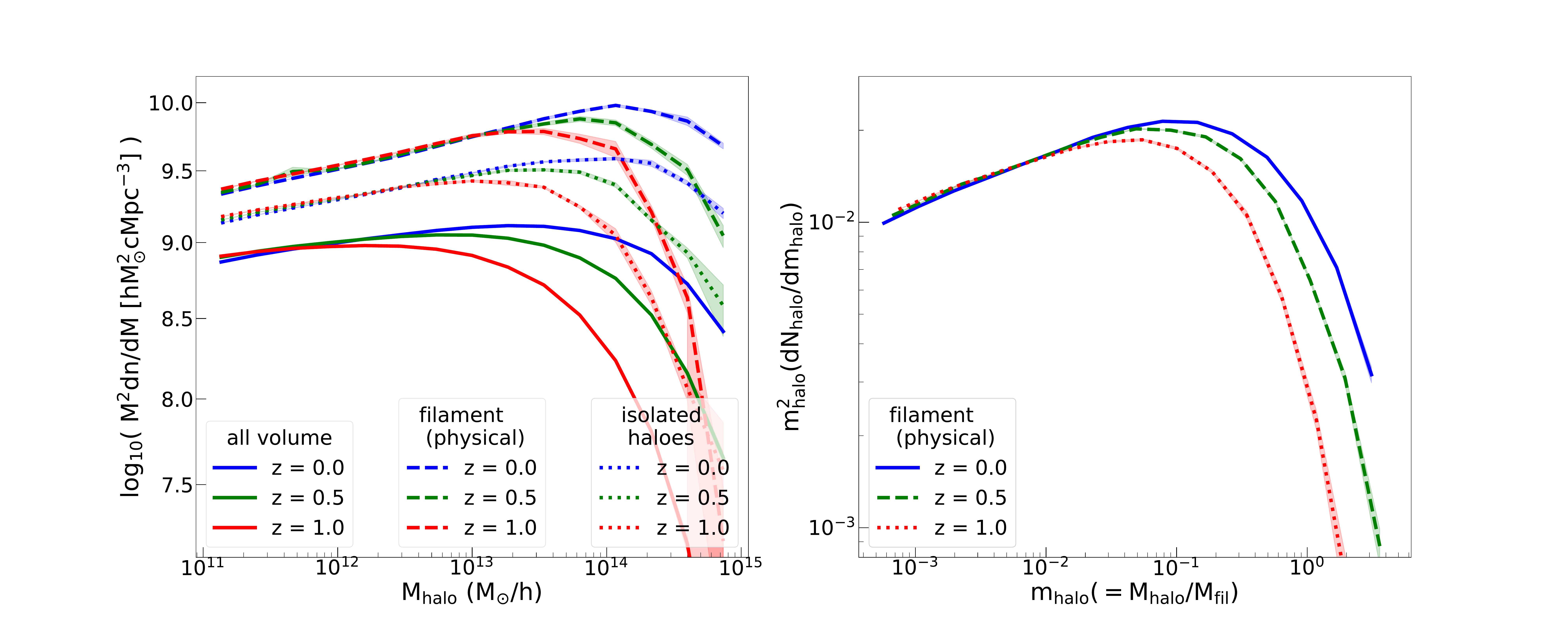}
    \end{minipage}%
\caption{Left panel: halo mass functions in different environments computed at different redshifts. Solid lines show the results from the whole volume and dashed lines show the results for filaments. In the latter case, haloes are counted in the filament region between two physically connected halo pairs, which is different from the definition of excess filament signal as shown in Figure \ref{params_fit_excluding_ends_all_fit} and \ref{rsep3d_binned_free_select}. For comparison, the dotted lines show the HMF around isolated haloes computed within the same filament volume. The boundary for each filament is given by the size-invariant $r_{\textrm{z}}/R_{\textrm{3D}}$ found in Figure \ref{rsep3d_binned_free_select}. Right panel: halo mass functions from the same filament environment but normalised by the total filament mass value between physically-connected halo pairs. The shaded region on each curve shows the Poisson error around the average determined from eight independent sub-boxes.}
\label{HMF_in_filament}
\end{figure*}

To characterise the halo distribution within filaments and see how it evolves with redshift, the left panel of Figure \ref{HMF_in_filament} shows the comoving halo mass function (HMF) computed for all volume or in the environment of filaments. Here the filament refers to the total mass between two physically connected halo pairs without the subtraction of matter distribution from the non-physical pairs, which is different from the excess mass $\Delta M_{\textrm{fil}}$. The mean total filament mass $M_{\textrm{fil}}$ is $2.36\times10^{14} ~h^{-1}M_{\odot}$ at $z = 0.0$, $2.04\times10^{14} ~h^{-1}M_{\odot}$ at $z = 0.49$ and $1.83\times10^{14} ~h^{-1}M_{\odot}$ at $z = 1.0$. For each filament, we choose the maximum cylindrical radius to be 0.35$R_{\textrm{3D}}$, which is a good approximation to the asymptotic value of the ratio $r_z/R_{\textrm{3D}}$ ratio for halo particles only from the lower middle panel in Figure \ref{rsep3d_binned_free_select} and the length of the filament, $L_{\textrm{fil}}$, excluding the halos at both ends. We note that $r_z/R_{\textrm{3D}}$ is approximately a constant and does not significantly evolve with redshift, and so the haloes are counted within this boundary at all redshifts. For comparison, on the same figure, we also include the HMF around isolated haloes, which characterises the contribution in proximity of isolated haloes within the same filament cylinder. It can be noticed that there is a small difference in amplitude between the HMF for physical filament and isolated haloes , which is the excess signal studied in this work.

The HMFs for the whole volume and filaments have a similar shape in that both have a exponential cut-off at some characteristic halo mass. However, the characteristic masses of this break are greater in the filaments than in the whole volume. For example, at $z=0$, HMFs computed in the filament environment peak at $M_{\textrm{vir}}\sim 10^{14} ~h^{-1}M_{\odot}$, which is similar to  the average $M_{\textrm{fil}}$. In contrast, the HMF in the simulation as a whole peaks at a lower mass of $M_{\textrm{vir}}\sim 10^{13.5} ~h^{-1}M_{\odot}$. At higher redshifts, the peak masses in both filaments and the entire simulation shift to lower halo masses.

The right panel shows the normalised HMFs for filaments, which is in bins of the ratio $m_{\textrm{halo}} = M_{\textrm{halo}}/M_{\textrm{fil}}$. This shows that stacked filaments are dominated by haloes with mass of $0.1$ -- $0.5 M_{\textrm{fil}}$, which is consistent with the result found in the left panel.

\section{Density profile and scaling relationships of filaments with selection effects}\label{sec::Density_profile_and scaling relationship_of_filaments_with_selection_biases}

In most observational studies, filament members between halo pairs are identified based on their separation in redshift and projected location on the sky. In this section, we fit Equation \ref{eqn::fil_density_profile} to groups of filaments selected by different criteria, and then compare the fits and filament properties to those obtained from previous section. We would like to investigate how the adopted selection criteria bias the results, and which criteria might be preferred. In Section \ref{ssec:Pair_Selection_in_Redshift_and_Projected_Distance_Space} we demonstrate how to find halo pairs in redshift and projected distance across simulation box. In Section \ref{ssec:Scaling_of_filament_properties_under_different_selection_cuts} we make comparisons of the fits and filament properties obtained under different selection cuts.

\subsection{Pair selection in redshift and projected separation}\label{ssec:Pair_Selection_in_Redshift_and_Projected_Distance_Space}

To be consistent with previous observational WL studies, we consider three cases with different separations in redshift and on the sky, with labels in the format  $R_{\textrm{2D}} \textrm{~range}/\Delta{z_{\textrm{sep}}}$. 

\begin{itemize}
    \item \textbf{6-10/0.002}: filaments between halo pairs selected with redshift separation $\Delta{z_{\textrm{sep}}}<0.002$ and projected 2D separation $6h^{-1} \textrm{cMpc} \leqslant R_{\textrm{2D}} \leqslant 10 h^{-1} \textrm{cMpc}$ \citep[e.g.][]{Seth_filament_paper, Tianyi_filament_paper};
    \item  \textbf{6-10/0.004}:  filaments between pairs with $\Delta{z_{\textrm{sep}}}<0.004$ and $6h^{-1} \textrm{cMpc} \leqslant R_{\textrm{2D}} \leqslant 10 h^{-1} \textrm{cMpc}$ \citep[e.g.][]{Clampitt_2014};
    \item \textbf{3-5/0.002}:  filaments between pairs with $\Delta{z_{\textrm{sep}}}<0.002$ and $3h^{-1} \textrm{cMpc} \leqslant R_{\textrm{2D}} \leqslant 5 h^{-1} \textrm{cMpc}$ \citep[e.g.][]{new_filament_paper}.
\end{itemize}

To identify these in the simulation, we must calculate the redshifts of the haloes. We treat the Cartesian $Z$ axis as the line-of-sight direction, therefore the projected distance separations between haloes are simply calculated as $\sqrt{\Delta X^2 + \Delta Y^2}$. The difference in redshifts between two haloes depends on the Hubble expansion and the peculiar motion of haloes along the line of sight. The redshift of an individual halo is
\begin{equation}\label{eqn::redshift_calculation}
    1+z_{\textrm{halo}} = (1+\overline{z})(1+z_{\textrm{p}}),
\end{equation}
where $\overline{z}$ is given by
\begin{equation}\label{eqn::z_bar_calculation}
    \overline{z} = z_{\textrm{sim}}+\frac{100h_{0}\Delta{\textrm{Z}}\sqrt{\Omega_{m,0}(1+z_{\textrm{sim}})^3 + (1-\Omega_{m,0})}}{c},
\end{equation}
and $z_{\textrm{p}}$ is given by
\begin{equation}\label{eqn::z_p_calculation}
    z_{\textrm{p}} = \frac{v_z}{c}\,
\end{equation}
where $v_z$ is the halo's peculiar velocity along the line-of-sight (Z-direction) in $\textrm{km/s}$, $z_{\textrm{sim}}$ is the redshift given by the snapshot. $\Delta{\textrm{Z}}$ is the line-of-sight comoving separation between the halo position and an arbitrary reference point in units of $h^{-1}$ cMpc. In this study, the reference point is fixed to be 500 $h^{-1} \textrm{cMpc}$. We shall investigate the impact of various chosen $\Delta z$ on the resulting filament properties in next section.

\subsection{Scaling of the filament properties and density parameters under different selection cuts }\label{ssec:Scaling_of_filament_properties_under_different_selection_cuts}

\begin{figure*}
    \begin{minipage}[htb!]{0.9\textwidth}
        \centering
        \includegraphics[width=\linewidth]{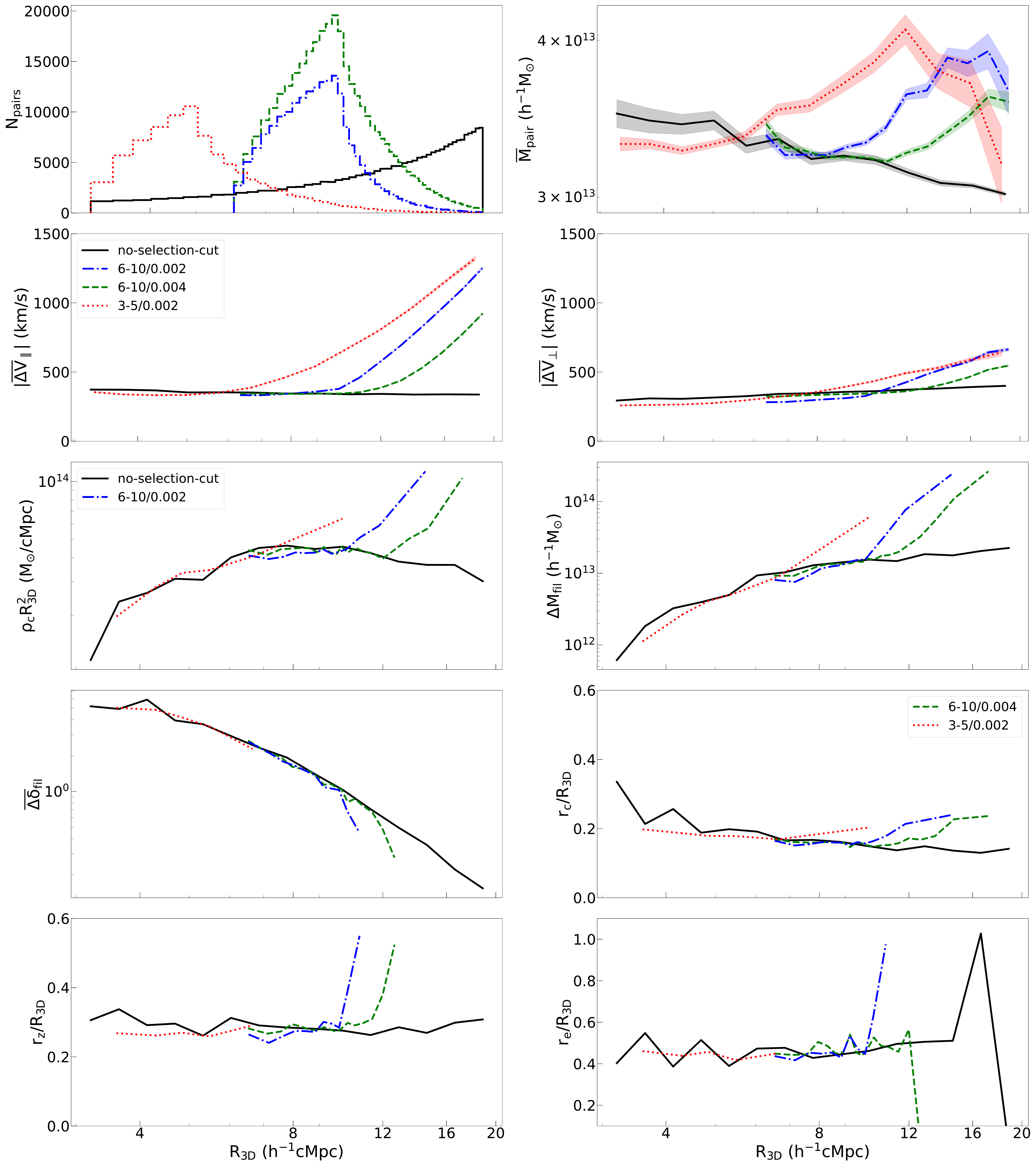}
    \end{minipage}%
\caption{The evolution of some filament-related properties with $R_{\textrm{3D}}$ under different selection criteria: no-selection-cut case (black solid line); 6-10/0.004 case (green dashed line); 6-10/0.002 case (blue dash-dotted line); 3-5/0.002 case (red dotted line). For illustration, only results obtained at snapshot $z = 0.49$ are shown here. First row: the pair distribution as a function of $R_{\textrm{3D}}$ for different selection criteria (\textit{left}); virial mass of pairs as a function of $R_{\textrm{3D}}$ (\textit{right}); Second row: relative velocity difference between two halo ends along the cylinder axis as a function of $R_{\textrm{3D}}$ (\textit{left}); relative velocity difference between two halo ends perpendicular to the cylinder axis as a function of $R_{\textrm{3D}}$ (\textit{right}). Shaded areas indicate the standard deviation around the mean. Third to fifth row: best fitting values of filament density profile obtained under different selection criteria with the same definition as Figure \ref{rsep3d_binned_free_select}. The full scaling of the best fitting density profile parameters when selection cuts applied is shown in Figure \ref{rsep3d_binned_selection_different_selection}.}
\label{rsep3d_binned_selection_bias_comparison}
\end{figure*}

To understand how selection effects may bias the selection of pairs, and hence the filament properties, we first focus on the properties of the halo pairs under different selection criteria based on their projected distance and redshift separations. The first two rows in Figure \ref{rsep3d_binned_selection_bias_comparison} shows filament properties as a function of $R_{\textrm{3D}}$ with different selection criteria. Only results obtained at the snapshot $z = 0.49$ are shown for illustration, although the results from $z = 0.0$ and $z = 1.03$ have consistent trends. Four different properties are considered here. The pair distribution as a function of $R\sbr{3D}$ is shown on the left panel in first row. It is noticeable that there is a sharp cutoff for long filaments when different selection criteria are applied. In the same row, we show the pair virial mass averaged per $R_{\textrm{3D}}$ bin, $\overline{M}_{\textrm{pair}}$, as a function of the filament length on the right. It can be seen that, when filaments are purely identified based on their 3D positions (using the criteria discussed in Section \ref{sec::fil_identify}, denoted as the ``no-selection-cut'' case), $\overline{M}_{\textrm{pair}}$ is decreasing as the  filament length increases. However, with the selection applied there is a tendency to select long filaments that are connected by massive haloes and are themselves more massive. This behaviour can also be seen from the $\Delta M_{\textrm{fil}}$ panel. This quantity is shown in the third row of Figure \ref{rsep3d_binned_selection_bias_comparison}, where for comparison, the best fitting parameters as a function of $R_{\textrm{3D}}$ for cases with different selection cuts applied are over-plotted. Mass for large-sized filaments are biased to a much higher values compared to the mass value obtained with the ``no-selection-cut'' case at the same $R_{\textrm{3D}}$ bin.
The second row of Figure \ref{rsep3d_binned_selection_bias_comparison} show the relative velocity between two halo ends projected along or perpendicular to the filament cylinder axis, denoted as $|\overline{\Delta V}_{\parallel}|$ (left) and $|\overline{\Delta V}_{\bot}|$ (right). $|\overline{\Delta V}_{\parallel}|$ signifies the relative motion of halo ends along the axis: a larger value of $|\overline{\Delta V}_{\parallel}|$ means that two halo ends have a larger relative infall, while $|\overline{\Delta V}_{\bot}|$ signifies the transverse relative velocity perpendicular to the filament axis.  

This behaviour can be understood as follows. The selection criteria include a maximum redshift separation. We interpret these biases as a tendency to select massive halo pairs that are aligned along the line of sight such that their high pairwise infall allow these halo pairs to satisfy the selection criteria. This is confirmed by comparing the 6-10/0.002 and 6-10/0.004 cases, for which at a fixed $R_{\textrm{3D}} \sim 12$ cMpc/$h$, the 6-10/0.002 selection has a higher mean pair mass compared to the 6-10/0.004 case. As an example, consider a filament of this length oriented along the line of sight; pure Hubble flow would yield $\delta z \sim 0.004$. However, if the halo pairs and filament are massive, they will have significant peculiar velocities toward each other: with an infall of $\sim 600$ \kms, the redshift separation is reduced to $\delta z \sim 0.002$ allowing them to satisfy the tighter redshift cut. 
This indicates that results from the ``6-10/0.004'' selection should be less biased compared to the other two criteria, and this is recommended for filament identification in future observational studies.

Properties of stacked filaments obtained under different selection criteria are summarised in Table \ref{table_pair_summary}, which includes the total size of each catalogue (number of halo pairs), the mean pair virial mass averaged over all filaments in the sample ($\langle {M}_{\textrm{pair}} \rangle$), best fitting parameters in the proposed density profile ($\langle \delta_{\textrm{c}} \rangle, \langle r_{\textrm{c}} \rangle, \langle r_{\textrm{e}} \rangle, \langle r_{\textrm{z}} \rangle$), the 3D excess filament mass. Errors in the best fitting parameters are obtained by bootstrapping all pairs as explained in Section \ref{ssec:Scaling_of_the_filament_density_parameters_no_selection_cut}, while errors in $\langle {M}_{\textrm{pair}} \rangle$ are simply Poisson errors around the mean.

\section{Discussion}\label{sec::discussion}
\subsection{Comparison with previous observational results}

It is worth noting that, when projected, our proposed profile predicts a negative foreground and background due to the fact that our computed stacked excess density profile goes to negative values at large radii (see Figure \ref{params_fit_excluding_ends_all_fit}). However, our N-body data only extend to 20 $h^{-1} \textrm{cMpc}$, we are uncertain whether our profile is still a reasonable fit when going to larger radii. The projected mass is, however, sensitive to the projection of the negative part. Therefore, the computation of the 2D projected excess filament masses by projecting Equation $\ref{eqn::fil_density_profile}$ along the line of sight is unreliable. As a possible extension of this work, it might be interesting to devise a hybrid profile that extrapolates Equation \ref{eqn::fil_density_profile} to the quasi-linear outskirts, where perturbation theory is expected to be valid, or simply come up with an alternate fitting formula for the simulated filaments in projection.

Instead, we compare our predicted central excess mass density $\langle \delta_{\textrm{c}} \rangle$ from simulation to other observational values. Using deprojected weak gravitational lensing, \citet{Tianyi_filament_paper} measured a central excess mass density of $\delta_{\textrm{c}} = (9.23\pm3.80)$ for filaments at mean redshift $z=0.44$ between LRG pairs with selection criteria of $6h^{-1} \textrm{cMpc} \leqslant R_{\textrm{2D}} \leqslant 10 h^{-1} \textrm{cMpc}$ and $|\Delta{z}|\leqslant0.002$, assuming $h$ = 0.70. Our result from the simulation using the same selection criteria applied around the same redshift is comparable to the observation within 1-$\sigma$ uncertainty: we predict a $ \langle \delta_{\textrm{c}} \rangle = 6.69\pm 0.07$.

By stacking around one million CMASS LRG pairs and using CMB lensing, \citet{Degraff_fil} estimated the central excess mass density of $\delta_{\textrm{c}} = (4.5\pm1.9)$ for filaments at median redshift $z\sim0.55$, where their pairs are identified with $\Delta{z_{\textrm{sep}}}<0.002$ and $6h^{-1} \textrm{cMpc} \leqslant R_{\textrm{2D}} \leqslant 14 h^{-1} \textrm{cMpc}$. It should be noted that they assume a different functional form, and so the agreement when extrapolating to the central value may be different. In spite of these differences, our model predicts a central excess mass density of $\langle \delta_{\textrm{c}} \rangle = 6.7\pm 0.1$, consistent with their measurement.

\citet{new_filament_paper} stacked the lensing signal generated by 11,706 LOWZ LRG pairs, with criteria $\Delta{z_{\textrm{sep}}}<0.002$ and $3h^{-1} \textrm{cMpc} \leqslant R_{\textrm{2D}} \leqslant 5 h^{-1} \textrm{cMpc}$. They found $\delta_{\textrm{c}} = 14.1\pm3.1$ which is consistent with our estimate ($13.8\pm 0.2$).

It is also interesting to compare our models with measurements of the gas temperature and density, as probed by the thermal Sunyaev-Zel'dovich (tSZ) effect. The tSZ signal from the stacked filaments between LOWZ LRG pairs \citep{Tanimura_fil} yields $\delta\sbr{c} \times (T\sbr{e}/10^7 \textrm{K}) \times (r\sbr{c} / 0.5 h^{-1} \textrm{cMpc}) = 2.7 \pm 0.5$ for filaments at median redshift $z\sim0.3$, where here $\delta\sbr{c}$ and $r\sbr{c}$ refer to parameters of the electron distribution and $T\sbr{e}$ is the electron temperature. Their pairs are selected with $\Delta{z_{\textrm{sep}}}<0.002$ and $6h^{-1} \textrm{cMpc} \leqslant R_{\textrm{2D}} \leqslant 10 h^{-1} \textrm{cMpc}$. For the same selection criteria, we find $\langle r_c \rangle = 1.5 h^{-1}$ cMpc and $\langle \delta_{\textrm{c}} \rangle = 7.2$. If we assume that the electron distribution follows the dark matter distribution, this allows us to solve for the electron temperature, yielding  $T\sbr{e} = 1.2\times10^6$ K ($\sim$ 0.1 keV), which is close to the density-weighted electron temperature of the universe estimated by \citet{The_Cosmic_Thermal_History_SZ_Tomography} at the same redshift.

Finally, the halo mass function allows us to predict the stellar mass content of filaments. The stellar-to-halo mass relation (SHMR) gives the stellar content of haloes of a given mass, and can be obtained via abundance matching \citep[e.g.][]{Behroozi_SMHM_redshift_scaling} or weak gravitational lensing \citep[e.g.][]{HudsonGillisCoupon2015}. The SHMR may depend on the large-scale environment, i.e.\ it may be different in filaments compared to the SHMR of the Universe as a whole \citep[e.g.][]{galaxy_formation_in_outside_fil,galaxy_formation_envir_dependence}. While this remains an open question, it is certainly true that the abundance of haloes does depend on environment, as discussed in Section \ref{ssec:halo_mass_function} \citep[see also][]{Halo_abundances_within_the_cosmic_web,Tracing_cosmic_web}. This, by itself, will introduce a variation of the ratio of stellar mass to total mass in different environments. We will calculate the filament stellar content directly using the halo mass function for filaments and assuming a universal stellar mass-halo mass relation fit to galaxies observed in all environments.  The halo mass functions for filaments are directly measured in simulations, which are shown as dotted lines in the left panel of Figure \ref{HMF_in_filament}. To compute the predicted total stellar mass, we use the SHMR from \citet{SMHM_ratio_kravtsov} after accounting for the scatter between $M_{\ast}$ and $M_{\textrm{vir}}$. The advantage of the \citet{SMHM_ratio_kravtsov} approach is that they consider sum of the stars in central galaxies and satellites, as opposed to most other studies which only consider the stellar content of the central galaxy. This study adopted the same functional form as in \citet{Behroozi_SMHM_redshift_scaling}, who fit the SHMR of central galaxies only but at multiple redshifts. We therefore use low redshift fits from \citet{SMHM_ratio_kravtsov} but evolve them using the redshift scaling of \citet{Behroozi_SMHM_redshift_scaling}. The resulting predictions for $M_{\textrm{stellar}}/M_{\textrm{total}}$ in filaments are 0.76\% at $z$ = 0.0, 0.88\% at $z$ = 0.49 and 1.02\% at $z$ = 1.032. From observational studies of the stellar mass content in filaments, \citet{Tianyi_filament_paper} found $(0.60\pm0.30)$\% for CMASS galaxies at $z = 0.53$ and $(0.84\pm0.47)$\% for LOWZ galaxies at $z = 0.33$, which in good agreement with the values derived from simulations and the assumption that the stellar-to-halo mass relation is universal.

\begin{table*}
\centering
\caption{Summary of the the properties of filaments identified between pairs of haloes, where $\langle \ldots \rangle$ denotes an average of over all filaments in the sample. (1) mean mass of halo pairs in units of $10^{13}h^{-1} M_{\odot}$. (2) excess filament mass in units of $10^{13}h^{-1} M_{\odot}$, obtained by integrating Equation \ref{eqn::fil_density_profile} using the best fitting parameters from the last four columns in a cylinder with a radius of $r_{\textrm{z}}$ and a length of $L_{\rm fil}\equiv{R}_{\textrm{3D}} - 2R_{\textrm{vir, halo 1}} - 2R_{\textrm{vir, halo 2}}$. Columns (3) (4),(5) \& (6) are best fitting parameters obtained by fitting the over-density profile averaged over all filaments. $\langle \delta_{\textrm{c}} \rangle$ is computed by $\langle \rho_{\textrm{c}} \rangle/\rho\sbr{m}$. The radii $\langle r_{\textrm{c}} \rangle$, $\langle r_{\textrm{z}} \rangle$ and $\langle r_{\textrm{e}} \rangle$ are in units of $h^{-1} \textrm{cMpc}$.}
\begin{tabular}{c|cccccccc} 
 \hline
 Selection & $z_{\textrm{snapshot}}$  & Pair number &
 \vtop{\hbox{\strut $\langle M _{\textrm{pair}}\rangle$}\hbox{\strut (1)}} & 
 \vtop{\hbox{\strut $\langle \Delta M_{\textrm{fil}}\rangle$}\hbox{\strut (2)}}&
 \vtop{\hbox{\strut $\langle \delta_{\textrm{c}}\rangle$}\hbox{\strut (3)}}&
 \vtop{\hbox{\strut $\langle r_{\textrm{c}}\rangle$}\hbox{\strut (4)}}&
 \vtop{\hbox{\strut $\langle r_{\textrm{z}}\rangle$}\hbox{\strut (5)}}&
 \vtop{\hbox{\strut 
 $\langle r_{\textrm{e}}\rangle$}\hbox{\strut (6)}} \\ [0.5ex] 
 \hline\hline
  & 0.00 &335,560 & 4.00$\pm$0.01 & 2.11$\pm$0.04&4.29$\pm$0.07&1.46$\pm$0.02&3.21$\pm$0.04&6.63$\pm$0.28 \\ 
  1/10 of Total Sample (no-selection-cut)& 0.49 &210,667&  3.12$\pm$0.01& 1.48$\pm$0.03&3.67$\pm$0.06&1.35$\pm$0.02&2.90$\pm$0.04&5.67$\pm$0.18 \\ 
  & 1.03 &84,006 &  2.46$\pm$0.01 & 1.11$\pm$0.03&3.79$\pm$0.10&1.15$\pm$0.02&2.57$\pm$0.05&5.37$\pm$0.24 \\ 
 \hline
 All Pairs with& 0.00 &381,202 & 4.33$\pm$0.01&2.50$\pm$0.03&7.72$\pm$0.08&1.51$\pm$0.01&3.56$\pm$0.04&5.70$\pm$0.17 \\ 
  $6h^{-1} \textrm{cMpc} \leqslant R_{\textrm{2D}} \leqslant 10 h^{-1} \textrm{cMpc}$& 0.49 &203,455 &3.35$\pm$0.01  &1.79$\pm$0.02&6.69$\pm$0.07&1.41$\pm$0.02&3.20$\pm$0.04&5.04$\pm$0.13 \\
 $|\Delta{z}|\leqslant0.002$& 1.03 &65,394 & 2.59$\pm$0.01&1.34$\pm$0.02&6.45$\pm$0.10&1.25$\pm$0.02&2.95$\pm$0.05&4.95$\pm$0.18 \\
 \hline
 All Pairs with& 0.00 &631,006 & 4.19$\pm$0.01&2.47$\pm$0.03&6.20$\pm$0.05&1.56$\pm$0.01&3.40$\pm$0.03&5.60$\pm$0.11 \\ 
  $6h^{-1} \textrm{cMpc} \leqslant R_{\textrm{2D}} \leqslant 10 h^{-1} \textrm{cMpc}$& 0.49 &344,651 &3.27$\pm$0.01  &1.82$\pm$0.02&5.82$\pm$0.05&1.44$\pm$0.01&3.17$\pm$0.03&5.17$\pm$0.10 \\
 $|\Delta{z}|\leqslant0.004$& 1.03 &115,047 & 2.56$\pm$0.01&1.36$\pm$0.02&5.80$\pm$0.08&1.29$\pm$0.02&2.94$\pm$0.04&4.95$\pm$0.14 \\
 \hline
 All Pairs with& 0.00 &159,173 &4.48$\pm$0.02& 0.77$\pm$0.02&13.68$\pm$0.24&0.99$\pm$0.02&1.68$\pm$0.02&2.76$\pm$0.06 \\
 $3h^{-1} \textrm{cMpc} \leqslant R_{\textrm{2D}} \leqslant 5 h^{-1} \textrm{cMpc}$ & 0.49 &92,127 &3.40$\pm$0.01&0.56$\pm$0.01   &13.98$\pm$0.25&0.89$\pm$0.01&1.45$\pm$0.02&2.32$\pm$0.05  \\
 $|\Delta{z}|\leqslant0.002$& 1.03 &33,921 &2.63$\pm$0.01&0.45$\pm$0.01&15.91$\pm$0.36&0.78$\pm$0.01&1.31$\pm$0.02&2.11$\pm$0.05 \\
 \hline
\end{tabular}
\label{table_pair_summary}
\end{table*}

\begin{figure*}
    \begin{minipage}[htb!]{\textwidth}
        \centering
        \includegraphics[width=\linewidth]{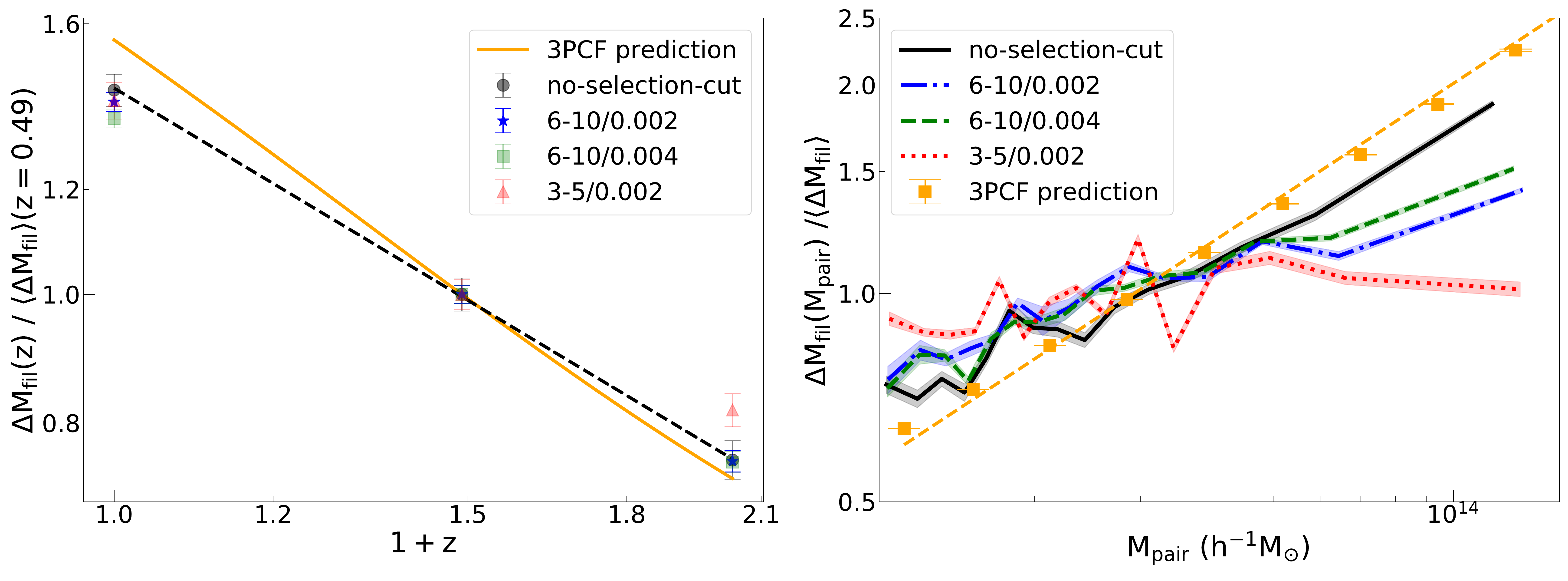}
    \end{minipage}%
\caption{Comparison between the filament scaling computed from simulations and theoretical predictions. Left panel: $\Delta M_{\textrm{fil}}$ as a function of redshift. Orange line shows the scaling of excess filament mass as redshift predicted by three-point correlation function (3PCF), while different data points show the $\Delta M_{\textrm{fil}}$ values under different selection criteria. Black dashed line shows the best-fit power law from the no-selection-cut case. Right panel: $\Delta M_{\textrm{fil}}$ as a function of pair virial mass for different selection criteria. For illustration, only results obtained at $z = 0.49$ are shown here. Squares indicate the prediction from the 3PCF, overplotted with the best-fit power-law (orange dashed line). Curves obtained at different selection criteria are differentiated by colors and styles with the same scheme as those in Figure \ref{rsep3d_binned_selection_bias_comparison}. Mass values are normalised by the $\langle \Delta M_{\textrm{fil}} \rangle$ values averaged over all halo pairs at $z = 0.49$, which are tabulated in Table \ref{table_pair_summary}.}
\label{mvir_mfil_all_selection}
\end{figure*}

\subsection{Comparison of filament scaling relations with predictions of perturbation theory}

In this section, we investigate how well the filament scaling relations are consistent with those based on theoretical predictions from perturbation theory. To model the excess mass between halo pairs, consider the galaxy-galaxy-matter three-point correlation function, which predicts that the redshift evolution of filament mass should depend on the halo bias $b(z)$ and the linear growth factor, $D(z)$, as follows \citep{Clampitt_2014}
\begin{equation}\label{eqn::3PCF_mass_cal}
    \Delta M_{\textrm{fil}} (z) = \left(\frac{b_1(z)}{b_1(z=0)}\right) \left(\frac{b_2(z)}{b_2(z=0)}\right) \left(\frac{D(z)}{D(z=0)}\right)^4 \Delta M_{\textrm{fil}}(z=0) \,,
\end{equation} 
where $b_1$ and $b_2$ are the linear bias factors of the haloes at the two ends of the filament. For the case where we are considering the ensemble average over all haloes above a threshold mass, then $b_1$ and $b_2$ can be replaced with the average number-weighted bias
\begin{equation}\label{eqn::bias_cal_num_weighted}
    \overline{b}(z) = \frac{\int_{10^{13} ~h^{-1}M_{\odot}}^{+\infty}\textrm{d}M\frac{\textrm{d}n}{\textrm{d}M}(M,z)b(M,z)}{\int_{10^{13} ~h^{-1}M_{\odot}}^{+\infty}\textrm{d}M\frac{\textrm{d}n}{\textrm{d}M}(M,z)}\,.
\end{equation}
Note that this differs from Equation \ref{eqn::bias_cal}, which is the \emph{mass-weighted} halo bias.

We can compare our filament mass measured from simulations at different redshifts to the 3PCF-predicted scaling. The scaling of excess filament mass, $\Delta M_{\textrm{fil}}$, as a function of redshift, as shown in the left panel of Figure \ref{mvir_mfil_all_selection}. The best-fit curve measured from the no-selection-cut sample follows a power law
\begin{equation}\label{eqn::Mfil_redshift_evol}
    \frac{\langle\Delta M\sbr{fil}\rangle(z)}{\langle\Delta M\sbr{fil}\rangle(z = 0.49)} = \left(\frac{1+z}{1.49}\right)^{-0.908\pm0.012}\,,
\end{equation}
where $\langle\Delta M\sbr{fil}\rangle(z = 0.49)$ is the excess filament mass measured at $z = 0.49$ in the unit of $h^{-1} M_{\odot}$ (see Table \ref{table_pair_summary}). The slope of this best fitting curve agrees to 8\% ($\sim 10\%$) with predicted 3PCF scaling. 

Another prediction of Equation \ref{eqn::3PCF_mass_cal} is that the filament mass should scale with the product of the bias of the haloes at either end of the filament. The resulting curves computed at $z = 0.49$ are shown in the right panel of Figure \ref{mvir_mfil_all_selection}. To obtain the curves as displayed in the panel, we first rank the pairs based on their $M_{\textrm{pair}}$ values, where $M_{\textrm{pair}}$ is simply the arithmetic mean of virial masses of the two haloes at either end of the filament. Then we group the sorted pairs into several bins such that the numbers of pairs per $M_{\textrm{pair}}$ bin are identical. The stacked over-density profile per $M_{\textrm{pair}}$ bin is fitted with Equation \ref{eqn::fil_density_profile} and the $\Delta M_{\textrm{fil}}$ is then computed by integrating the best fitting profile. The 3PCF-predicted values are obtained by using Equation \ref{eqn::3PCF_mass_cal}, where for each halo pair, $b(z)^2$ is expanded as $b(M_{\textrm{vir, halo 1}}, z)b(M_{\textrm{vir, halo 2}}, z)$. Squares represent the predicted $\Delta M_{\textrm{fil}}$ averaged per $M_{\textrm{pair}}$ bin. It can be seen that the overall trend between the squared points from theory and the curves from simulations are consistent. Approximated as a power law over the range $10^{13} h^{-1} M_{\odot} < M\sbr{pair} < 10^{14} h^{-1} M_{\odot}$, the no-selection-cut sample yields
\begin{equation}\label{eqn::Mfil_mpair_evol}
\frac{\Delta M\sbr{fil}(M\sbr{pair})}
{\langle\Delta M\sbr{fil}\rangle} = \left(\frac{M\sbr{pair}}{\langle M\sbr{pair}\rangle}\right)^{0.405\pm0.022}
\end{equation}
at $z = 0.49$, where $\langle\Delta M\sbr{fil}\rangle$, the ensemble average over all values of $M\sbr{pair}$ and $\langle M\sbr{pair}\rangle$ are given in Table \ref{table_pair_summary}. However, the 3PCF theory predicts a somewhat steeper mass dependence, with a power-law index of 0.57 (steeper by $\sim 30\%$). Both comparisons illustrated here suggest that in order to predict filament properties more precisely, one might need to include higher order corrections in the three-point correlation prediction.

\section{Conclusion}\label{sec::conclusion}

We have demonstrated that, similar to dark matter haloes, it is possible to describe the dark matter distribution of stacked filaments with a universal functional form. To achieve this, we proposed a four-parameter density profile and fit this to the  stacked profiles of filaments identified from the MultiDark Planck 2 simulation at three different redshifts ($z = 0.0, 0.49, 1.032$). Our profile provides reasonable fits out to distance of $ \sim 20 h^{-1} \textrm{cMpc}$ normal to the filament axis. We showed that the density profile parameters, $r\sbr{c}$ (core radius), $r\sbr{z}$ (zero-crossing radius) and $r\sbr{e}$ (``environmental'' radius), do not strongly evolve with redshift and approximately scale linearly with the filament 3D length.

We further studied how different observational selection criteria affect the scaling of filament density profile parameters as a function of filament length and redshift. We found that a choice of smaller redshift separation for long filaments would preferentially select more massive filaments, as the redshift separation of the pair is decreased by the infalling peculiar velocities. This issue can be mitigated by increasing the redshift separation. For example, an optimal choice could be $\Delta{z_{\textrm{sep}}}<0.004$ and $6h^{-1} \textrm{cMpc} \leqslant R_{\textrm{2D}} \leqslant 10 h^{-1} \textrm{cMpc}$. 

Our stacked filament masses and central density contrasts, $\langle \delta\sbr{c} \rangle$, are consistent with published observational studies. The dependence of filament mass on redshift and bias factors of the two haloes are similar to, but slightly shallower (by 10\% and 30\% respectively) than the predictions from perturbation theory for 3-point correlation function. With the universal profile and the scaling with halo mass and redshift, this allows one to construct a template filament profile for comparison with other observables such weak lensing, X-ray and radio data, and the Sunyaev-Zel'dovich effect.

We note that our proposed profile is calibrated against a DM-only simulation. It would be interesting to extend this analysis to universal profiles for stacked filaments in hydrodynamic simulations that include the physics of galaxy formation, and examine whether a similar functional form can provide a faithful description of matter distribution. 

Observationally, stacked filaments can be measured with the combination of an imaging survey (for weak lensing) and a redshift survey (to identify the halo pairs). In the near future, the Ultraviolet Near-Infrared Optical Northern Survey\footnote{https://www.skysurvey.cc/} \citep[UNIONS, see e.g.][]
{IbataMcConnachieCuillandre2017}, which overlaps large spectroscopic surveys in the North, will have sufficient precision for a detailed comparison of filament properties with our model. Future space-based missions such as Euclid, the Nancy Grace Roman Space Telescope, and the proposed Astrophysics Telescope for Large Area Spectroscopy \cite[ATLAS,][]{WangRobbertoDickinson2019}, which will collect redshifts in the same sky area covered by weak lensing observations from Roman, will measure filament profiles at higher redshifts and with still greater precision. These are important next steps to untangle the complex web of physical processes that impact the structure of our cosmic web.

\section*{Acknowledgements}

MH and NA acknowledge their respective NSERC Discovery grants. This work was supported by the University of Waterloo, Natural Sciences and Engineering Research Council of Canada (NSERC), and the Perimeter Institute for Theoretical Physics. Research at the Perimeter Institute is supported by the Government of Canada through Industry Canada, and by the Province of Ontario through the Ministry of Research and Innovation.
The CosmoSim database used in this paper is a service by the Leibniz-Institute for Astrophysics Potsdam (AIP).The MultiDark database was developed in cooperation with the Spanish MultiDark Consolider Project CSD2009-00064.
The authors gratefully acknowledge the Gauss Centre for Supercomputing e.V. (www.gauss-centre.eu) and the Partnership for Advanced Supercomputing in Europe (PRACE, www.prace-ri.eu) for funding the MultiDark simulation project by providing computing time on the GCS Supercomputer SuperMUC at Leibniz Supercomputing Centre (LRZ, www.lrz.de).

\section*{Data availability}
The raw MDPL2 simulation data and halo catalogues used in this paper are available at https://www.cosmosim.org/cms/simulations/mdpl2/. The filament data will be made available on request to the lead author.

\bibliographystyle{mnras}
\bibliography{draft}

\appendix
\onecolumn
\section{}
In this appendix, we include the data used for plotting Figure \ref{rsep3d_binned_free_select}. The best fitting normalised scale radii as well as the derived $\overline{\Delta \delta}_{\textrm{fil}}$ are provided in Table \ref{table_params_summary_selection_free}. $\overline{\Delta \delta}_{\textrm{fil}}$ is computed with $\Delta\rho_{\textrm{fil}}/\rho_{\textrm{bg}}$, where $\rho_{\textrm{bg}}$ is the background matter density either using all particles in the simulation box, or just the particles in haloes with $M_{\textrm{vir}} \geq 10^{11} h^{-1} M_{\odot}$. $\rho_{\textrm{fil}}$ is determined by the best fitting integrated mass (as shown in the top middle panel of Figure \ref{rsep3d_binned_free_select}) divided by the filament cylinder volume. These data can be interpolated to a different filament sizes for the study of filament properties.
\begin{table*}
\centering
\caption{Table of the best fitting density profile parameters for the no-selection-cut case, as plotted in Figure \ref{rsep3d_binned_free_select}}.
\begin{tabular}{|c|c|cccc|cccc|}
\hline
&&\multicolumn{4}{c|}{all particles} &\multicolumn{4}{c|}{halo particles only} \\
 $z_{\textrm{snapshot}}$ & ${R}_{\textrm{3D}}$ & $\overline{\Delta \delta}_{\textrm{fil}}$& $r_{\textrm{c}}/{R}_{\textrm{3D}}$&  $r_{\textrm{z}}/{R}_{\textrm{3D}}$& $r_{\textrm{e}}/{R}_{\textrm{3D}}$& $\overline{\Delta \delta}_{\textrm{fil}}$& $r_{\textrm{c}}/{R}_{\textrm{3D}}$&  $r_{\textrm{z}}/{R}_{\textrm{3D}}$& $r_{\textrm{e}}/{R}_{\textrm{3D}}$ \\ [0.5ex] 
  \hline
  &3.20&5.00$\pm0.83$&0.23$\pm0.08$&0.30$\pm0.02$&0.51$\pm0.05$&7.95$\pm3.02$&0.23$\pm0.17$&0.26$\pm0.05$&0.42$\pm1.32$\\ 
  &3.64&3.75$\pm0.58$&0.21$\pm0.04$&0.31$\pm0.03$&0.60$\pm0.08$&5.10$\pm1.85$&0.19$\pm0.10$&0.32$\pm0.06$&0.66$\pm1.46$\\ 
  &4.13&5.65$\pm0.58$&0.22$\pm0.05$&0.32$\pm0.02$&0.50$\pm0.05$&10.32$\pm1.97$&0.20$\pm0.10$&0.32$\pm0.03$&0.50$\pm0.10$\\ 
  &4.69&4.83$\pm0.39$&0.23$\pm0.02$&0.34$\pm0.02$&0.55$\pm0.05$&7.34$\pm1.20$&0.23$\pm0.03$&0.38$\pm0.03$&0.61$\pm0.10$\\
  &5.32&3.13$\pm0.23$&0.21$\pm0.02$&0.35$\pm0.02$&0.61$\pm0.05$&4.67$\pm0.67$&0.20$\pm0.02$&0.38$\pm0.03$&0.71$\pm0.10$\\
  &6.04&3.06$\pm0.22$&0.18$\pm0.01$&0.32$\pm0.01$&0.56$\pm0.05$&4.74$\pm0.75$&0.17$\pm0.02$&0.35$\pm0.03$&0.62$\pm0.49$\\
  &6.85&2.55$\pm0.16$&0.18$\pm0.01$&0.30$\pm0.01$&0.51$\pm0.04$&3.77$\pm0.49$&0.17$\pm0.02$&0.34$\pm0.02$&0.58$\pm0.08$\\
  0.00&7.77&2.23$\pm0.13$&0.17$\pm0.01$&0.30$\pm0.01$&0.48$\pm0.03$&3.55$\pm0.44$&0.17$\pm0.01$&0.33$\pm0.02$&0.49$\pm0.07$\\
  &8.82&1.45$\pm0.10$&0.16$\pm0.01$&0.31$\pm0.01$&0.52$\pm0.05$&1.86$\pm0.28$&0.16$\pm0.01$&0.37$\pm0.02$&0.69$\pm1.29$\\
  &10.01&1.15$\pm0.07$&0.15$\pm0.01$&0.29$\pm0.01$&0.50$\pm0.04$&1.42$\pm0.19$&0.14$\pm0.01$&0.35$\pm0.02$&0.78$\pm1.76$\\
  &11.36&0.91$\pm0.05$&0.15$\pm0.01$&0.29$\pm0.01$&0.49$\pm0.04$&1.15$\pm0.14$&0.14$\pm0.01$&0.34$\pm0.02$&0.72$\pm1.43$\\
  &12.89&0.62$\pm0.03$&0.15$\pm0.01$&0.30$\pm0.01$&0.56$\pm0.08$&0.84$\pm0.09$&0.15$\pm0.01$&0.34$\pm0.02$&0.76$\pm0.95$\\
  &14.64&0.42$\pm0.02$&0.14$\pm0.01$&0.31$\pm0.01$&0.72$\pm0.71$&0.53$\pm0.07$&0.14$\pm0.01$&0.37$\pm0.02$&$\infty$\\
  &16.61&0.32$\pm0.02$&0.13$\pm0.01$&0.28$\pm0.01$&0.72$\pm0.73$&0.41$\pm0.04$&0.13$\pm0.01$&0.32$\pm0.02$&$\infty$\\
  &18.86&0.19$\pm0.01$&0.14$\pm0.01$&0.31$\pm0.01$&$\infty$&0.24$\pm0.04$&0.15$\pm0.01$&0.38$\pm0.02$&$\infty$\\
 \hline
  &3.20&5.20$\pm0.78$&0.34$\pm0.19$&0.31$\pm0.03$&0.40$\pm0.09$&11.00$\pm2.99$&0.50$\pm0.30$&0.30$\pm0.12$&0.36$\pm0.33$\\
  &3.64&4.94$\pm0.54$&0.21$\pm0.02$&0.34$\pm0.02$&0.55$\pm0.06$&8.97$\pm2.04$&0.21$\pm0.09$&0.39$\pm0.06$&0.57$\pm0.73$\\
  &4.13&5.91$\pm0.59$&0.26$\pm0.05$&0.29$\pm0.02$&0.39$\pm0.04$&13.11$\pm1.96$&0.26$\pm0.14$&0.27$\pm0.02$&0.32$\pm0.06$\\
  &4.69&3.95$\pm0.32$&0.19$\pm0.02$&0.30$\pm0.02$&0.51$\pm0.04$&7.26$\pm1.06$&0.18$\pm0.03$&0.31$\pm0.03$&0.51$\pm0.09$\\
  &5.32&3.68$\pm0.25$&0.20$\pm0.02$&0.26$\pm0.01$&0.39$\pm0.03$&6.24$\pm0.88$&0.19$\pm0.07$&0.27$\pm0.03$&0.39$\pm0.09$\\
  &6.04&2.92$\pm0.22$&0.19$\pm0.01$&0.31$\pm0.01$&0.47$\pm0.04$&5.18$\pm0.68$&0.18$\pm0.01$&0.35$\pm0.03$&0.53$\pm0.08$\\
  &6.85&2.34$\pm0.14$&0.17$\pm0.01$&0.29$\pm0.01$&0.48$\pm0.03$&3.24$\pm0.44$&0.15$\pm0.01$&0.35$\pm0.03$&0.68$\pm1.79$\\
  &7.77&1.93$\pm0.13$&0.17$\pm0.01$&0.28$\pm0.01$&0.43$\pm0.04$&2.96$\pm0.36$&0.16$\pm0.01$&0.34$\pm0.02$&0.54$\pm0.12$\\
  0.49&8.82&1.40$\pm0.08$&0.16$\pm0.01$&0.28$\pm0.01$&0.45$\pm0.03$&2.15$\pm0.22$&0.16$\pm0.01$&0.33$\pm0.02$&0.54$\pm0.10$\\
  &10.01&1.03$\pm0.06$&0.15$\pm0.01$&0.28$\pm0.01$&0.46$\pm0.03$&1.61$\pm0.18$&0.15$\pm0.01$&0.32$\pm0.02$&0.55$\pm0.47$\\
  &11.36&0.71$\pm0.04$&0.14$\pm0.01$&0.26$\pm0.01$&0.50$\pm0.04$&1.03$\pm0.10$&0.13$\pm0.01$&0.32$\pm0.02$&0.82$\pm1.56$\\
  &12.89&0.50$\pm0.03$&0.15$\pm0.01$&0.29$\pm0.01$&0.51$\pm0.05$&0.67$\pm0.07$&0.15$\pm0.01$&0.36$\pm0.02$&0.91$\pm1.07$\\
  &14.63&0.36$\pm0.02$&0.14$\pm0.01$&0.27$\pm0.01$&0.51$\pm0.08$&0.49$\pm0.05$&0.14$\pm0.01$&0.34$\pm0.02$&$\infty$\\
  &16.61&0.22$\pm0.02$&0.13$\pm0.01$&0.30$\pm0.01$&1.03$\pm0.77$&0.31$\pm0.05$&0.13$\pm0.01$&0.36$\pm0.03$&$\infty$\\
  &18.85&0.15$\pm0.01$&0.14$\pm0.01$&0.31$\pm0.01$&$\infty$&0.23$\pm0.03$&0.15$\pm0.01$&0.37$\pm0.02$&$\infty$\\
 \hline
  &3.21&7.77$\pm1.22$&0.20$\pm0.05$&0.29$\pm0.03$&0.41$\pm0.89$&21.95$\pm5.10$&0.18$\pm0.12$&0.30$\pm0.13$&0.35$\pm2.40$\\
  &3.63&7.13$\pm1.38$&0.21$\pm0.06$&0.31$\pm0.03$&0.41$\pm0.09$&21.30$\pm4.11$&0.22$\pm0.16$&0.31$\pm0.06$&0.28$\pm1.15$\\
  &4.13&4.74$\pm0.75$&0.20$\pm0.03$&0.28$\pm0.03$&0.42$\pm0.07$&11.22$\pm2.38$&0.18$\pm0.09$&0.30$\pm0.05$&0.43$\pm0.14$\\
  &4.68&5.48$\pm0.62$&0.24$\pm0.10$&0.27$\pm0.03$&0.31$\pm0.08$&11.68$\pm1.91$&0.20$\pm0.12$&0.30$\pm0.05$&0.35$\pm0.68$\\
  &5.31&4.32$\pm0.32$&0.16$\pm0.01$&0.26$\pm0.01$&0.39$\pm0.04$&9.05$\pm1.38$&0.13$\pm0.03$&0.27$\pm0.04$&0.43$\pm0.14$\\
  &6.03&3.03$\pm0.21$&0.16$\pm0.01$&0.27$\pm0.01$&0.43$\pm0.03$&5.97$\pm0.70$&0.15$\pm0.02$&0.32$\pm0.03$&0.51$\pm0.82$\\
  &6.86&2.14$\pm0.16$&0.17$\pm0.01$&0.28$\pm0.01$&0.44$\pm0.04$&4.40$\pm0.52$&0.18$\pm0.03$&0.33$\pm0.03$&0.47$\pm0.09$\\
  1.03&7.78&1.77$\pm0.12$&0.15$\pm0.01$&0.25$\pm0.01$&0.41$\pm0.04$&3.10$\pm0.35$&0.15$\pm0.04$&0.30$\pm0.03$&0.51$\pm1.04$\\
  &8.81&1.13$\pm0.08$&0.14$\pm0.01$&0.27$\pm0.01$&0.48$\pm0.04$&2.18$\pm0.28$&0.15$\pm0.02$&0.33$\pm0.03$&0.56$\pm1.73$\\
  &10.01&0.94$\pm0.06$&0.14$\pm0.01$&0.25$\pm0.01$&0.43$\pm0.04$&1.76$\pm0.18$&0.15$\pm0.02$&0.30$\pm0.02$&0.52$\pm0.55$\\
  &11.36&0.60$\pm0.04$&0.14$\pm0.01$&0.27$\pm0.01$&0.49$\pm0.04$&0.96$\pm0.10$&0.16$\pm0.02$&0.35$\pm0.03$&0.66$\pm1.43$\\
  &12.89&0.42$\pm0.03$&0.13$\pm0.01$&0.27$\pm0.01$&0.56$\pm0.33$&0.62$\pm0.08$&0.14$\pm0.02$&0.36$\pm0.04$&$\infty$\\
  &14.63&0.30$\pm0.02$&0.15$\pm0.01$&0.27$\pm0.01$&0.43$\pm0.05$&0.48$\pm0.07$&0.16$\pm0.03$&0.36$\pm0.04$&0.61$\pm0.94$\\
  &16.61&0.20$\pm0.02$&0.14$\pm0.01$&0.28$\pm0.01$&0.70$\pm0.73$&0.32$\pm0.04$&0.15$\pm0.02$&0.38$\pm0.04$&$\infty$\\
  &18.85&0.11$\pm0.01$&0.11$\pm0.01$&0.28$\pm0.02$&$\infty$&0.13$\pm0.03$&0.12$\pm0.01$&0.43$\pm0.04$&$\infty$\\
 \hline
\end{tabular}
\label{table_params_summary_selection_free}
\end{table*}

\section{Scaling of the best fitting density profile parameters under different selection cuts }

\begin{figure*}
    \begin{minipage}[b]{0.45\textwidth}
        \centering
        \includegraphics[width=\linewidth]{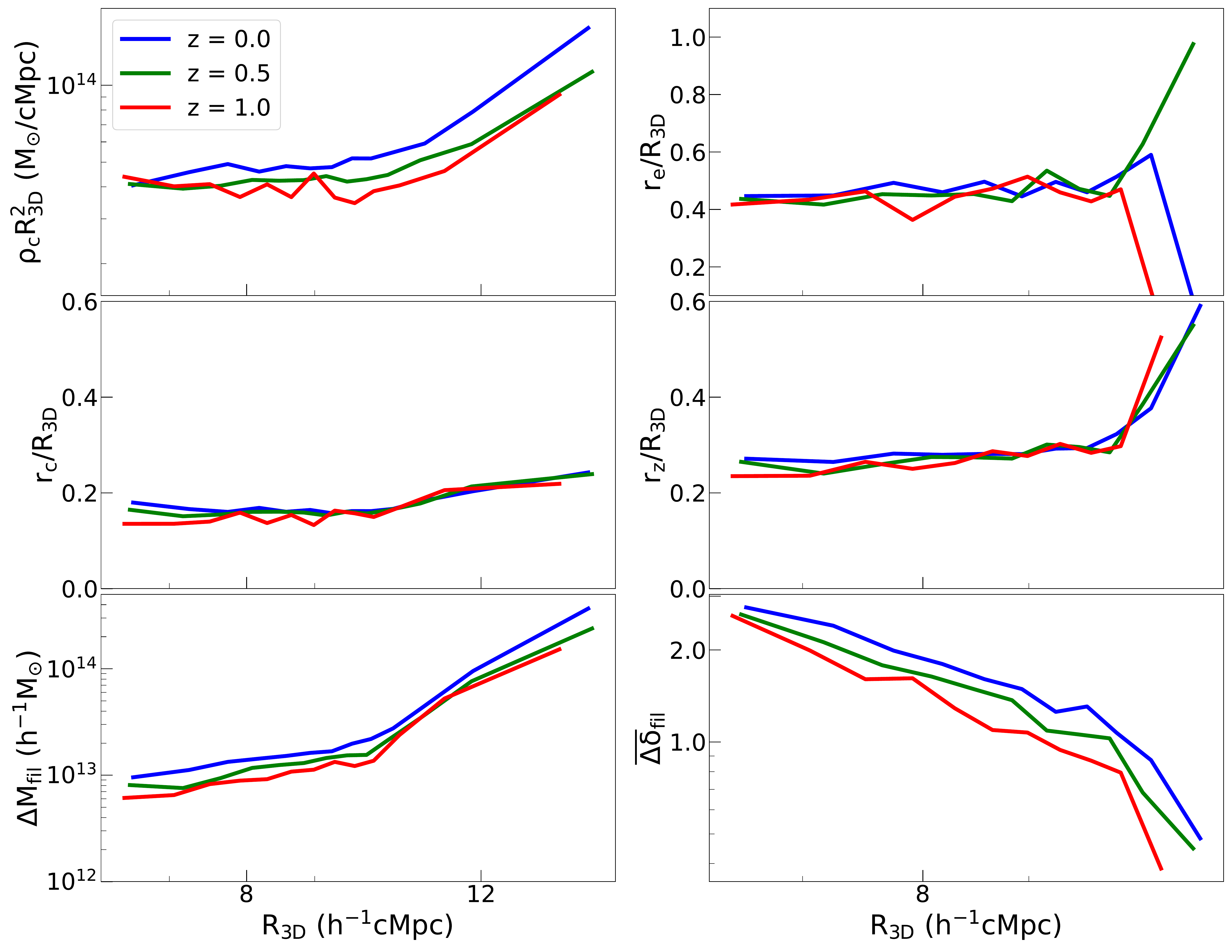}
    \end{minipage}
    \begin{minipage}[b]{0.45\textwidth}
        \centering
        \includegraphics[width=\linewidth]{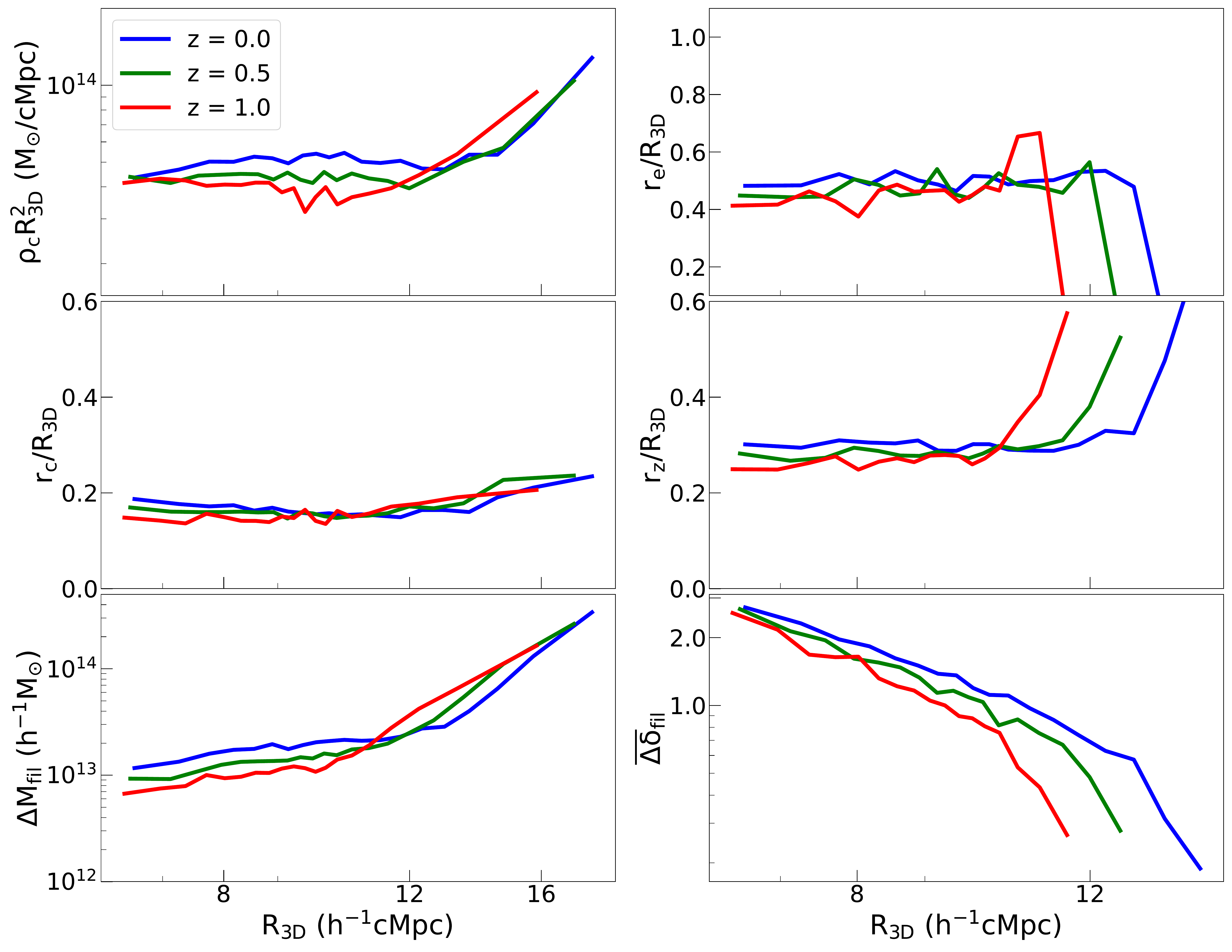}
    \end{minipage}
    \begin{minipage}[b]{0.45\textwidth}
        \centering
        \includegraphics[width=\linewidth]{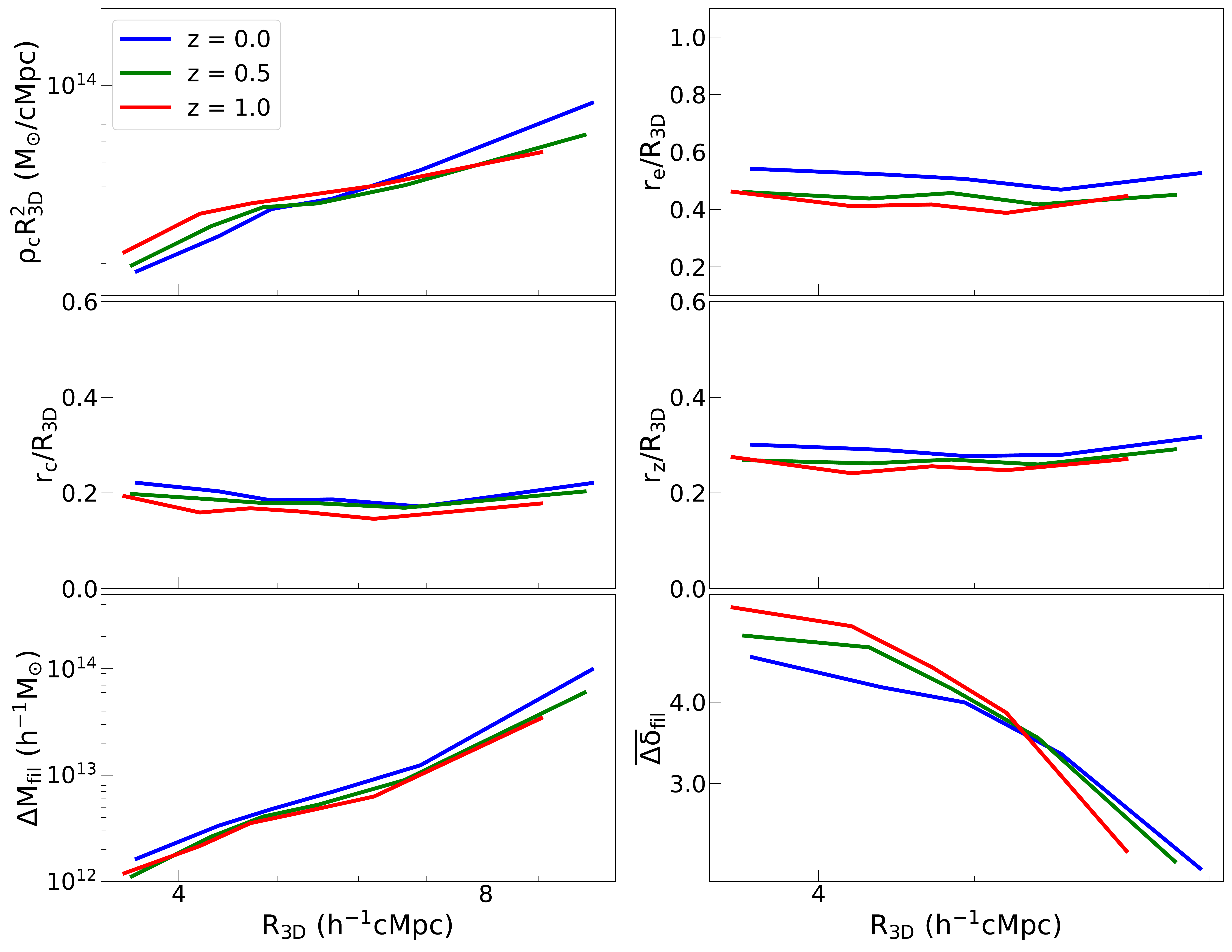}
    \end{minipage}%
\caption{The scaling of the best fitting density profile parameters under different selection criteria, with the same definition as Figure \ref{rsep3d_binned_free_select}. On each panel, redshift evolution of different quantities are distinguished by different colors. Top left: 6-10/0.002 case. Top right: 6-10/0.004 case. Bottom: 3-5/0.002 case.}
\label{rsep3d_binned_selection_different_selection}
\end{figure*}

Figure \ref{rsep3d_binned_selection_different_selection} shows the best fitting parameter as a function of $R_{\textrm{3D}}$ and redshift for cases with different selection cuts applied. The top left panel shows the fit from the ``6-10/0.002'' case and the top right panel shows the results from the ``6-10/0.004'' case, while the bottom panel shows the fits from the ``3-5/0.002'' case. 

Compared to the curves displayed in Figure \ref{rsep3d_binned_free_select}, in the bottom panel, one can see that when applying selection criteria, the best-fitting core density $\rho_{\textrm{c}}$ and the excess filament mass are increasing significantly for long filaments. This is consistent with the idea discussed in Section \ref{ssec:Scaling_of_filament_properties_under_different_selection_cuts} that, due to the selection of a small redshift separation, one preferentially selects massive filaments because these induce large infalling peculiar velocities of the two end halos, which then satisfy the narrow limits in redshift-space. This bias is reduced when the chosen value for redshift separation is larger. Again, this suggests that ``6-10/0.004'' is preferred while identifying filament members because not only does it improve the statistics by including more halo pairs, but it also yields less biased filament properties.

\label{lastpage}
\end{document}